\documentstyle[pre,twocolumn,aps,psfig]{revtex}
\begin{document}
\draft
\title{Fluid Models for Kinetic Effects on Coherent Nonlinear Alfv\'en
Waves. II. Numerical Solutions}
\author{M. V. Medvedev$^{1,}$
\thanks{ E-mail: {\tt mmedvedev@ucsd.edu}
Internet: {\tt http://sdphpd.ucsd.edu/\~{ }misha/mm.html} }$^,$
\thanks{ Also at the Institute for Nuclear Fusion, 
Russian Research Center ``Kurchatov Institute", Moscow 123182, Russia.}
, V. I. Shevchenko$^2$
, P. H. Diamond$^{1,}$ 
\thanks{ Also at General Atomics, San Diego, California 92122.}
, and V. L. Galinsky$^3$ }
\address{$^1$ Physics Department, University of California at San Diego,
La Jolla, California 92093-0319}
\address{$^2$ Electrical \& Computer Engineering Department,
University of California at San Diego, La Jolla, California 92093-0407}
\address{$^3$ Scripps Institution of Oceanography,
University of California at San Diego, La Jolla, California 92093-0210}

\author{
\parbox[t]{6in}{\small
The influence of various kinetic effects (e.g. Landau damping, 
diffusive and collisional dissipation, and finite Larmor radius terms) on the nonlinear
evolution of finite amplitude 
Alfv\'enic wave trains in a finite-$\beta$ environment is systematically 
investigated using a novel, kinetic nonlinear 
Schr\"odinger (KNLS) equation. The dynamics of Alfv\'en waves is sensitive
to  the sense of polarization as well as the angle of propagation with
respect to the ambient magnetic field. Numerical solution for the case with 
Landau damping reveals the formation of dissipative structures, 
which are quasi-stationary, S-polarized directional 
(and rotational) discontinuities which self-organize from parallel propagating, 
linearly polarized waves. Parallel propagating
circularly polarized packets evolve to a few circularly polarized 
Alfv\'en harmonics on large scales. Stationary arc-polarized rotational 
discontinuities form from obliquely propagating waves. Collisional 
dissipation, even if weak, introduces enhanced wave damping when
$\beta$ is very close to unity. Cyclotron motion
effects on resonant particle interactions introduce
cyclotron resonance into the nonlinear Alfv\'en wave dynamics.\newline
}
}
\maketitle

\section{Introduction}
\label{sec:intro}

The envelope dynamics of nonlinear Alfv\'en waves at small-$\beta$
are thought to be governed by the derivative nonlinear Schr\"odinger (DNLS) 
equation, which describes parametric coupling with acoustic modes
\cite{Cohen}. 
It is believed that the dynamics of such waves changes drastically in a
collisionless,  finite-$\beta$ plasma, due to several kinetic effects, 
especially collisionless (Landau) damping of ion-acoustic oscillations
\cite{Rogister,Mjolhus,Spangler}. In a highly dissipative regime, the
ion-acoustic {\em quasi-}mode is no longer a plasma eigenmode.
Thus, it is more natural (according Refs.\ \cite{Hollweg,Lee}) to view
the mechanism of Landau damping of Alfv\'en waves as  particle trapping
by both the magnetic mirror force and electric field. On account of the
inevitable mathematical difficulties of a kinetic approach, many 
numerical simulations have been performed. 
DNLS-based simulations with beam pumping (comet modeling)
\cite{Shevch1,Shevch2}, hybrid magnetohydrodynamic-particle simulations
\cite{Hada1,Hada2}, hybrid particle ions-fluid electrons simulations 
\cite{Vasquez1,Vasquez2,Vasquez3} have been performed.

In a previous paper \cite{us}, we systematically accounted for 
ion kinetic effects (but neglected finite conductivity, discussed elsewhere 
\cite{Mjolhus,conductivity2}) on parallel and slightly
oblique propagating waves in homogeneous plasma, to the order at 
which the DNLS is derived. A fluid representation for Landau damping
\cite{Hammet} allowed us to derive a relatively simple evolution equation
for finite amplitude Alfv\'en waves in finite-$\beta$ environments,
typical of the solar wind. Collisional damping was also included.
For slightly oblique propagation, the effects of finite Larmor radius result
in nonlinear dispersion-type 
integral terms. The terms which describe kinetic effects are usually integral 
(nonlocal) in nature, and reflect the finite ion transit time through the envelope
modulation of an Alfv\'en train. However, for the  weakly collisional (viscous)
dissipation and for the  small Larmor radius limits, these integral operators asymptote 
to (local) differential operators. This is not the case for Landau damping, since no
intrinsic scale exists. The envelope evolution equation obtained is thus  a nonlocal,
integro-differential equation which is not not amenable to 
{\em analytical} solution.

In this paper, we explore numerical solution of
the kinetically modified derivative nonlinear
Schr\"odinger (KNLS) equation. Some preliminary results of this
investigation are given
in Ref.\ \cite{Galinsky}. We again classify the solutions by the
kinetic effects which are taken into account. For simplicity, 
we usually consider parallel propagating waves, however the results are valid
for oblique waves since they still obey the KNLS equation 
\cite{oblique1,oblique2,us}.

We discover that (especially for the case of Landau damping) new
types of waveform emerge. The known observed ``final states'' of nonlinear 
wave steepening (i.e. shock structures) are classified as:
\begin{itemize}
\item[1.)] {\em collisional {\em (hydrodynamic)} shocks} for which 
nonlinear steepening is limited by {\em collisional} (viscous) dissipation,
which sinks energy from small scales (i.e. from high-$k$ harmonics);
\item[2.)] {\em collisionless shocks} (common in astrophysical plasma)
in which nonlinear steepening is limited by {\em dispersion} resulting in 
soliton-type structures with sufficient energy content in high-$k$
harmonics.
\end{itemize}
A new, large class of waveforms, {\em dissipative structures}, appears when
dissipation acts on all scales (recall Landau damping has no intrinsic 
characteristic scale). Thus, we add 
\begin{itemize}
\item[3.)] {\em dissipative structures}, for which nonlinear steepening 
is limited by {\em collisionless} (scale invariant) damping. Harmonic spectra
of such structures contain predominantly low-$k$ harmonics, i.e.
energy resides at small scales. Fast rotation of the
of wave magnetic field vector (i.e. rotational or directional discontinuity formation)
is another common feature of these structures.
\end{itemize}
Dissipative structures emerge in plasma with $T_i\sim T_e$ for
a wide range of $\beta$: $0.5\le\beta\le1.5$, which is typical
for the solar wind plasma. Thus, we argue that such structures are likely
a common constituent of solar wind Alfv\'enic turbulence. Note here that
the dissipative structures discussed here emerge without external energy 
source in the system. Thus, the amplitude of them decreases with time.

Numerous {\em in situ} observations of the solar wind magnetic activity
have revealed the nonlinear nature of MHD waves \cite{NLwaves1,NLwaves2}.
Recent observations indicate the existence
of directional (e.g. rotational) discontinuities, i.e.
regions of rapid phase jumps where the amplitude also varies 
\cite{NLwaves1,DD}. Several types of directional/rotational
discontinuities which might be distinguished by their phase portraits
have been observed. There are
(i) discontinuities of the ``S-type'', at which the magnetic field vector rotates
first through some angle (less than or close to $90^\circ$) in one direction,
followed by rotation in the opposite direction through an angle larger 
than $180^\circ$ (typically, $180^\circ<\Delta\phi\le270^\circ$) 
\cite{DD,S-type}, and (ii) arc-polarized
discontinuities, where the magnetic field vector rotates along an arc through
an angle less than $180^\circ$ \cite{NLwaves1,arc}.
At directional discontinuities, the fast phase jump is accompanied by
moderate amplitude modulation ($\delta B\sim B$). At the rotational 
discontinuities, the amplitude modulation is small or negligible
($\delta B\ll B$). It is shown in this
paper that all these commonly occurring wave structures are manifested in
the nonlinear evolution of Alfv\'en waves in finite $\beta$, isothermal plasma
for different initial conditions and propagation angle. 

The theory of nondissipative  Alfv\'en waves governed by the conservative
DNLS equation predicts nonlinear wave steepening and formation of
waveforms with steep fronts. These waveforms are unstable to
modulations. Thus, spiky, many-soliton structures are
emitted from the steep edge. These oscillatory, (roughly) circularly
polarized soliton-like structues interact nonlinearly to yield strongly
turbulent, spatially irregular wave profiles [see next section].
However, the DNLS theory fails
to explain the formation of rotational and directional discontinuities.
A further step was to include the linear damping of Alfv\'en waves due to 
finite plasma conductivity. It was shown that the nonlinear wave relaxes
to a shock train and constant-$B$  rotational discontinuities that rotate
the field through exactly $180^\circ$ \cite{train1,train2,Cohen}.
Inspite of this, the DNLS theory was unable to explain the existance
and dynamics of both (i) the S-polarized directional and 
rotational discontinuities and (ii) arc-polatized rotational discontinuities
with rotation of less than $180^\circ$. Recent particle code simulations
\cite{Vasquez1,Vasquez2,Vasquez3}
have shown that Landau damping may drastically change the dynamics of
Alfv\'en waves and result in the emergence of such wave structures.

In this paper, we use a recently developed \cite{us} analytical
model of the KNLS equation to investigate the influence of Landau damping 
and several other kinetic effects on the dynamics of Alfv\'en waves. 
The main claim of this paper is that {\em all} the discontinuous wave structures
discussed above are {\em distinct solutions} of the same {\em simple} analytical
model for different initial conditions, e.g. initial wave polarization
and wave propagation angle.
A (quasi-) parallel, initially circularly polarized, 
amplitude modulated wave evolves to a single, purely circular, lowest-$k$
harmonic decoupled from dissipation. A (quasi-) parallel linearly polarized
wave forms a {\em dissipative structure} which is characterized by S-type,
fast phase rotation and moderate amplitude variation, i.e. has the
requisite properties for classification as a localized, S-type directional/rotational
discontinuity. Such discontinuities are intrinsically {\em dissipative}, as their
amplitude continuously decreases with time. An obliquely propagating
wave forms an arc-type rotational discontinuity, which
does not experience even minimal dissipation. 
The effects of finite Larmor radius (significant at oblique angles)
are shown to tend to suppress arc-polarized structure formation. 
Low-$k$, finite-amplitude circular waves form instead. Collisional (diffusive)
dissipation of a ponderomotively driven acoustic mode is significant 
(even in a weakly collisional plasma) if the $\beta$ value is close to unity, thus
increasing plasma temperature in this  region.

The remainder of this paper is organized as follows. In Section\ 
\ref{sec:initial}, we discuss the method of calculation, initial conditions
and standard DNLS solutions (i.e. the limiting case of $\beta=0$). In Section\
\ref{sec:landau}, we solve the KNLS for the case of Landau damping.
In Section\ \ref{sec:viscous}, we investigate an effect of collisional 
damping. In Section\ \ref{sec:larmor}, the influence of finite Larmor 
radius corrections on the dynamics of slightly obliquely
propagating waves is investigated.
Section\ \ref{sec:concl} presents a summary of the result obtained and
some concluding remarks.

\section{Numerical method, initial conditions, and DNLS solutions}
\label{sec:initial}

The general KNLS equation may be written as 
\begin{equation}
\frac{\partial b}{\partial t_e}+\frac{1}{2}\frac{\partial}{\partial z}
\left(U_{NL}b\right)+i\frac{v_A^2}{2\Omega_i}
\frac{\partial^2b}{\partial z^2}=0 ,
\label{mainKNLS}
\end{equation}
where $b=b(z-v_At, t_e)=(b_x+ib_y)/B_0$ is the wave magnetic field,
$v_A$ and $\Omega_i$ are the Alfv\'en speed and proton ion-cyclotron 
frequency, and $t_e=(b^2/B_0^2)t$ represents slow envelope evolution. 
The nonlinear ponderomotive plasma velocity perturbation is
\begin{equation}
U_{NL}=\frac{v_A}{2}\left(c_1\left(|b|^2-\langle|b|^2\rangle\right)
+c_2\widehat{\cal K}\left[|b|^2-\langle|b|^2\rangle\right]\right) .
\label{u}
\end{equation}
Here $c_1$ and $c_2$ are functions of plasma parameters, i.e. $\beta$, 
$T_e/T_i$, etc., only. Throughout the paper, we set $T_e=T_i$ for
simplicity. $\widehat{\cal K}$ is an integral operator which represents 
kinetic effects. The detailed form of it depends on the 
effects  under consideration. We now introduce the dimensionless
coordinate and time respectively as $\zeta=z/\bar z$ and $\tau=t_e/\bar t$,
where $\bar z=50c/\omega_p$ and $\bar t=200/\Omega_i$. Here 
$\omega_p$ is the proton plasma frequency and $c$ is the speed of light. 
Then, writing $b(\zeta, \tau)=L^{-1/2}\sum_k b_k \exp(i\lambda_k\zeta)$,
where $\lambda_k=2\pi k/L$ ($L$ is the dimensionless domain length), the 
equation for $k$-th harmonic, $b_k$, follows as
\begin{eqnarray}
& &\frac{\partial b_k}{\partial\tau}+ i\eta\lambda_k^2b_k
 \label{knls-num}\\
& &{ }
+i\lambda_k\left\{c_1\left(|b|^2-\langle|b|^2\rangle\right)b
+c_2\widehat{\cal K}\left[|b|^2-\langle|b|^2\rangle\right]b\right\}_k=0 ,
\nonumber
\end{eqnarray}
where $\eta=0.04$ is the dispersion parameter. This equation was solved
for periodic boundary conditions using a predictor-corrector scheme and
a fast Fourier transform to calculate nonlinearities.

For $\beta=0$ no kinetics impacts the wave dynamics \cite{us}, so
Eq.\ (\ref{mainKNLS}) reduces to the familiar DNLS equation with $U_{NL}$
simply
\begin{equation}
U_{NL}=\frac{1}{2}\frac{v_A}{1-\beta}
\left(|b|^2-\langle|b|^2\rangle\right) .
\label{u-dnls}
\end{equation}
The DNLS is integrable and has the exact (soliton) solution 
\cite{soliton,WFlaM,M,conductivity2}. We pick a 
standing (in the wave frame) soliton solution
($b_0$ is the soliton amplitude):
\begin{mathletters}
\begin{eqnarray}
b(\zeta)&=&\frac{b_0\exp(i\Theta)}
{\cosh\left[\left(\zeta-L/2\right)b_0^2/2\eta\right]} ,
\label{solit-a}\\
\Theta(\zeta)&=&\frac{2}{3}\arctan
\left\{\sinh\left[\left(\zeta-L/2\right)b_0^2/2\eta\right]\right\} .
\label{solit-b}
\end{eqnarray}
\end{mathletters}
To avoid effects of periodic boundary
conditions, we have chosen the domain length $L$ equal to $32\pi$
and taken 8192 harmonics (and spatial points) so that 
$-4096\le n\le4096$. The test run has shown excellent agreement
with the analytical solution
during the time of computation (up to $\tau=40$, i.e. 8,000 cyclotron
periods).

\begin{figure}[p]
\begin{center}
\psfig{file=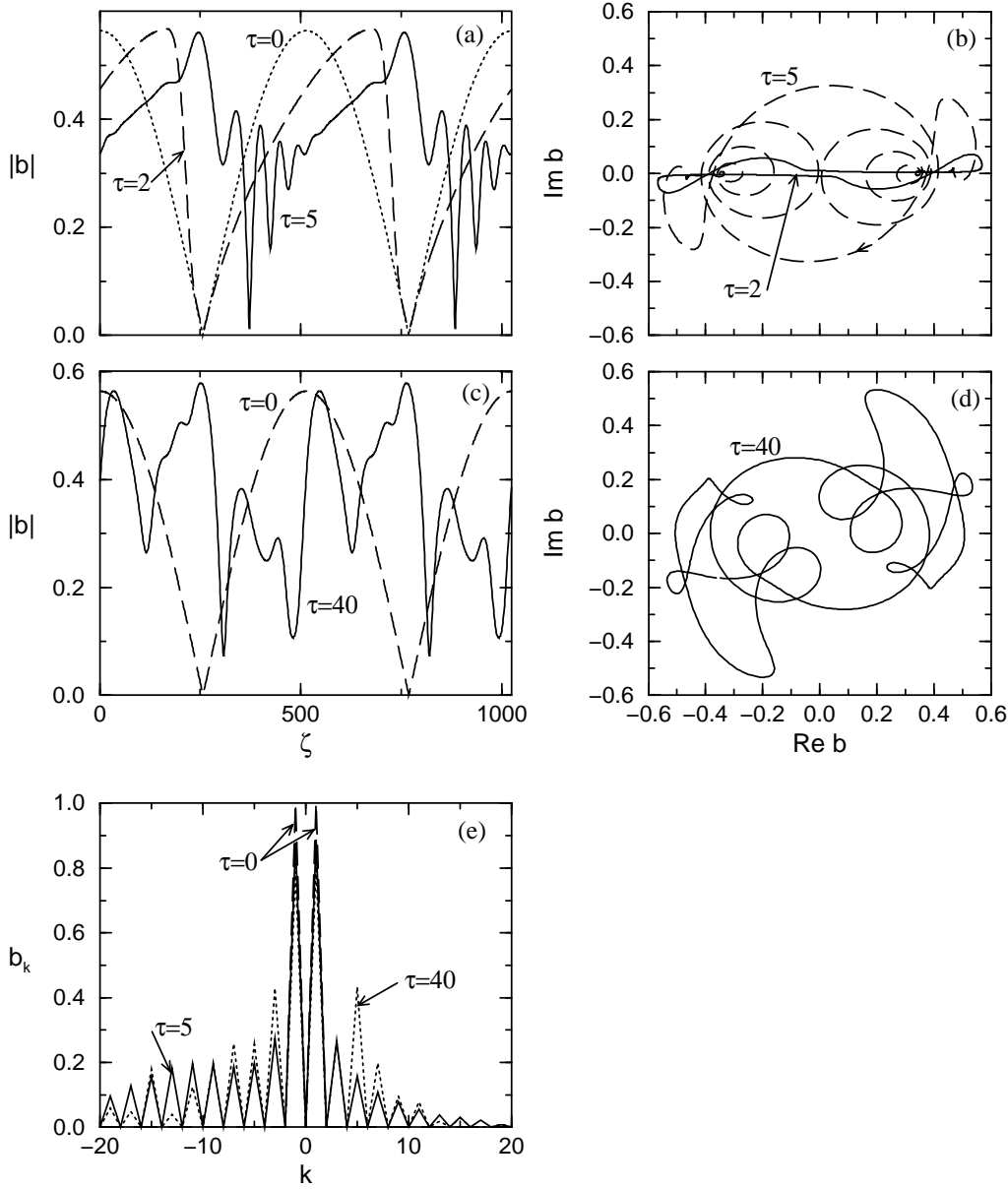,height=9.in}
\vskip-4cm
\caption{ Wave evolution from the DNLS with $\beta=0, M_1=1, M_2=0$ for
{\em linearly} polarized, sinusoidal wave initial condition. (a) and (c) - wave
profiles, (b) and (d) - phase diagrams (hodographs), (e) - harmonic
spectra at $\tau=0, 5, 40$. At $\tau=0$ only $k=1, -1$ harmonics are excited.
Note  an irregular wave structure at late times (i.e. $\tau=40$).}
\label{fig:lin}
\end{center}
\end{figure}

\begin{figure}[p]
\begin{center}
\psfig{file=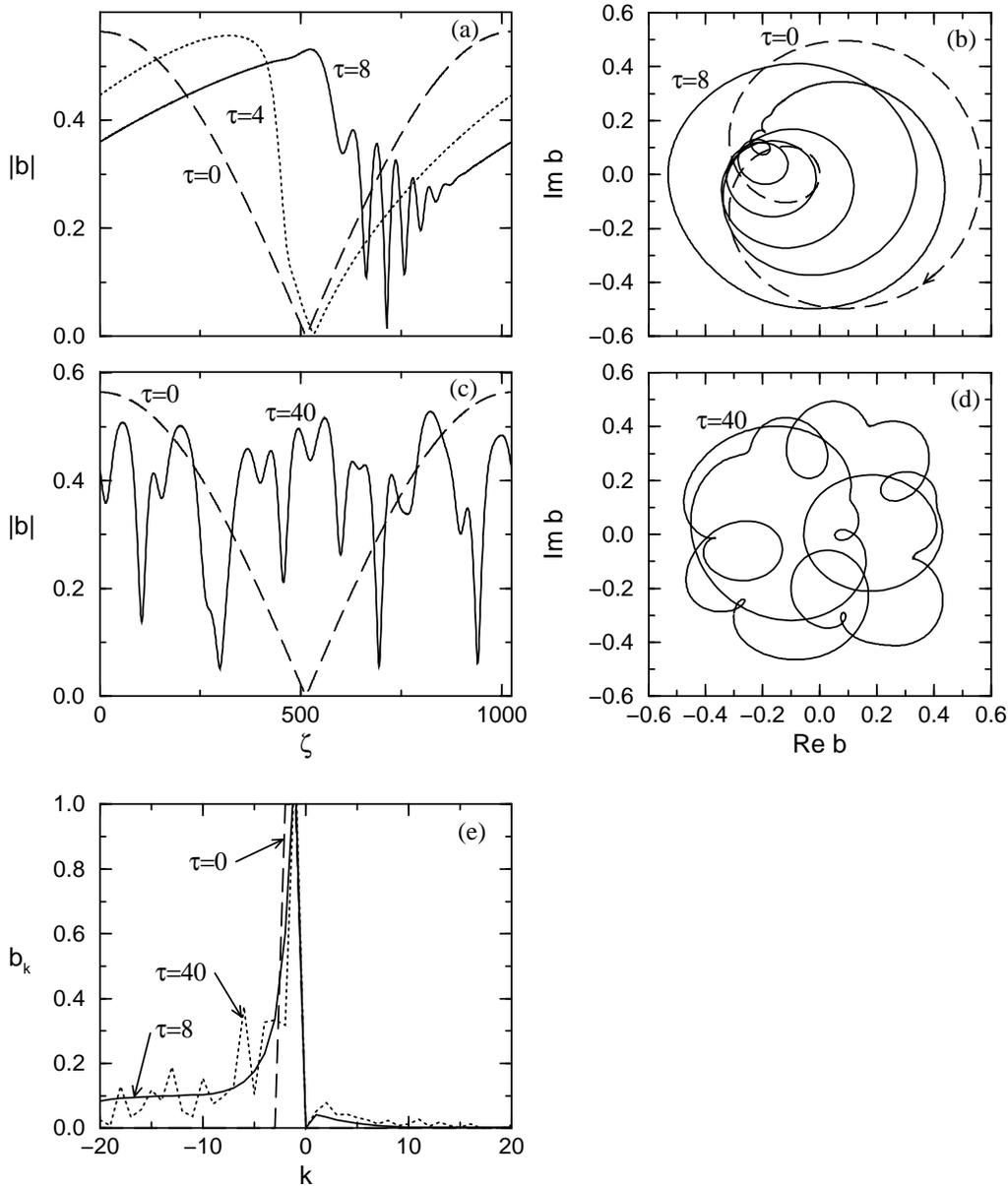,height=9.in}
\vskip-4cm
\caption{Wave evolution from the DNLS with $\beta=0$ for   
{\em circularly} polarized, sinusoidally amplitude modulated initial condition.
(a) and (c) - wave profiles, (b) and (d) - phase diagrams (hodographs), 
(e) - harmonic spectra at $\tau=0, 8, 40$. At $\tau=0$ only $k=-1, -2$ 
harmonics are excited. Note the irregular wave structure at late 
times $\tau=40$.}
\label{fig:circ}
\end{center}
\end{figure}

As high-amplitude magnetic perturbations in the solar wind probably evolve
from small-amplitude (linear) ones, it is reasonable to examine the
nonlinear evolution of finite-amplitude periodic waves of 
both linear and circular polarizations. In these cases, we have taken
$L=4\pi$ and 1024 harmonics and spatial points, i.e. $-512\le n\le512$.
Results of reference runs for amplitude modulated linear and circular
polarizations are shown in Figs.\ \ref{fig:lin} and \ref{fig:circ},
respectively. 
[Note here that only amplitude modulated waves experience
nonlinear evolution, a purely circularly polarized wave with 
$b\sim\exp(i\lambda\zeta)$ is an exact solution of the DNLS
(and KNLS) for which nonlinear effects do not enter since
$|b|^2=1$.] The initial wave profiles are given by two initially
excited Fourier harmonics. For linear polarizations, we pick
$b_{-1}=b_1=1$, all others are zeroes, for circular polarizations
we pick $b_{-2}=b_{-1}=1$, for elliptical polarizations, we  pick
$b_{-1}=1.1,~ b_{1}=0.9$. Thus, the waves are left-hand polarized.
However amplitude profiles, $|b|$ vs. $\zeta$, 
(Figs.\ \ref{fig:lin}a, \ref{fig:circ}a) look alike, their phase portraits
(Figs.\ \ref{fig:lin}b, \ref{fig:circ}b) and harmonic spectra
(Figs.\ \ref{fig:lin}e, \ref{fig:circ}e)  differ significantly.
All the waves exhibit  the nonlinear steepening phase of a front
at earlier times ($\tau\sim2$). Dispersion further limits steepening
and produces (at times $\tau\sim5$) small-scale parasitic,
oscillatory, circularly polarized (even for initial linear polarization)
wave structures. These oscillations are just DNLS solitons radiated
from the steep front of the wave train by modulation instability.
(By different authors, such a process is referred to as ``dispersive steepening''.) 
The direction of polarization is shown by an arrow
on graphs  \ref{fig:lin}b, \ref{fig:circ}b, indicating right hand polarization.
At later times, $\tau\sim40$, nonlinear processes result in 
a wave magnetic field which is completely irregular,
 as seen from Figs.\ \ref{fig:lin}c, \ref{fig:lin}d, \ref{fig:circ}c,
\ref{fig:circ}d, as well as from their harmonic spectra,
(Figs.\ \ref{fig:lin}e, \ref{fig:circ}e). In these cases strong, large-amplitude 
Alfv\'enic turbulence is developed.

\section{Collisionless dissipative Alfv\'en trains}
\label{sec:landau}
\subsection{Numerical solutions}

The KNLS equation with collisionless (Landau) dissipation 
\cite{Rogister,Mjolhus,Spangler,us,WFlaM,FlaMW}is Eq.\
(\ref{mainKNLS}), together with the nonlinear velocity perturbation written as
\begin{equation}
U_{NL}=\frac{v_A}{2}\left\{M_1\left(|b|^2-\langle|b|^2\rangle\right)
+M_2\widehat{\cal L}\left[|b|^2-\langle|b|^2\rangle\right]\right\} ,
\label{u-landau}
\end{equation}
where
\begin{eqnarray*}
M_1&=&{(1-\beta)+\widehat\chi^2_\|(1-\beta/\gamma)\over
 (1-\beta)^2+\widehat\chi^2_\|(1-\beta/\gamma)^2 } , \\
M_2&=&-\widehat\chi_\|\beta
{(\gamma-1)/\gamma\over (1-\beta)^2+\widehat\chi^2_\|(1-\beta/\gamma)^2 } ,
\end{eqnarray*}
$\gamma=3,~ c_s^2=\gamma T/m,~ \beta=c_s^2/v_A^2$, and 
$\widehat\chi_\|=(8\beta/\pi\gamma)^{1/2}$ is the parallel heat flux 
dissipation coefficient which models the intrinsically kinetic effects of 
resonant particles. Here $\widehat{\cal L}$ is the resonant particle integral 
(Hilbert) operator:
\begin{equation}
\widehat{\cal L}[f](x)=\frac{1}{\pi}\int_{-\infty}^\infty
\frac{\cal P}{x'-x}f(x') {\rm d}x'\doteq i\frac{k}{|k|}f_k .
\end{equation}
The sign $\doteq$ denotes equivalence to a Fourier image and ${\cal P}$
denotes a principal value integral.

\begin{figure}[p]
\begin{center}
\psfig{file=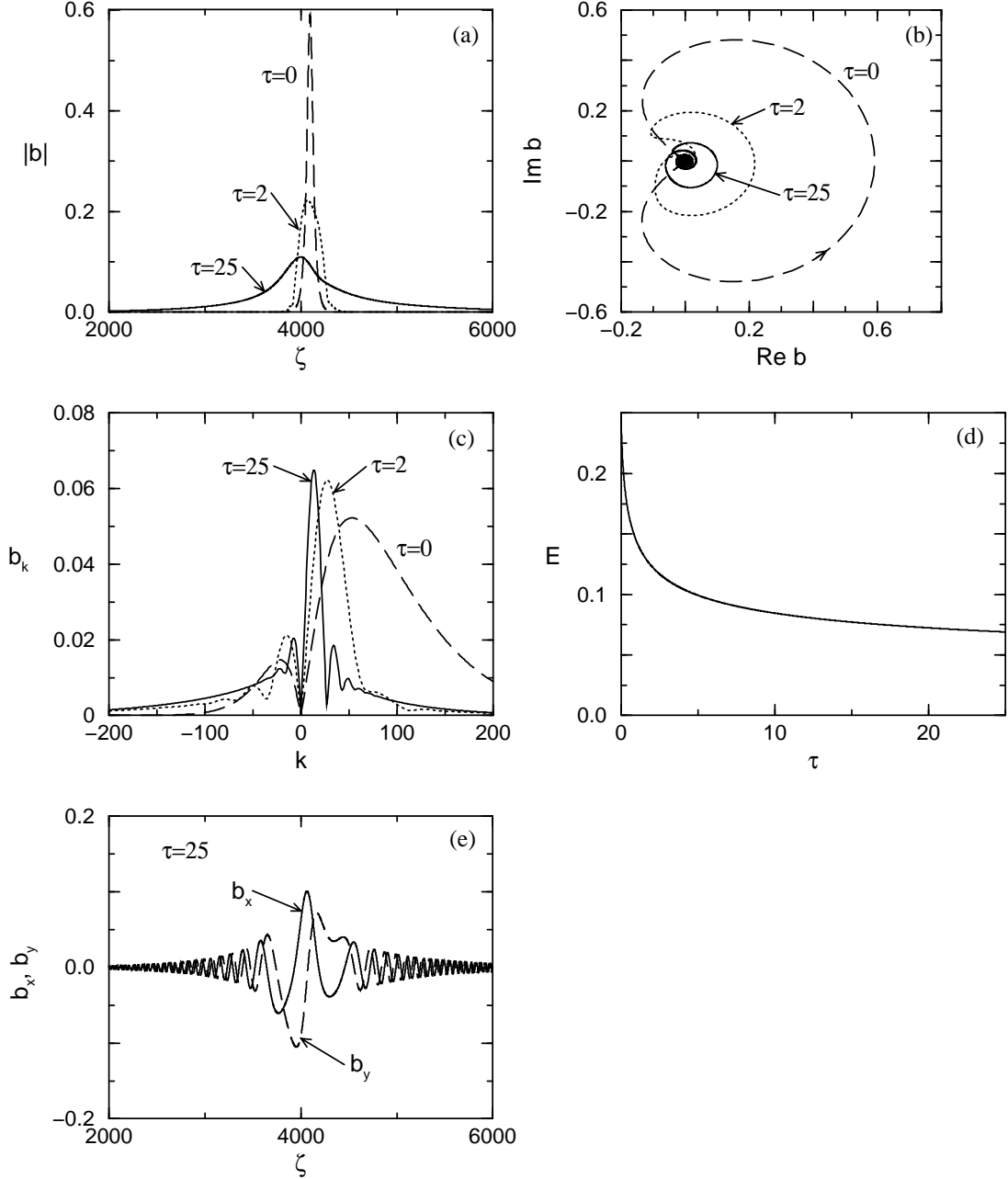,height=9.in}
\vskip-4cm
\caption{Evolution for a parallel propagating, 
{\em soliton} initial condition for the KNLS with 
Landau damping for $\beta=1$, $M_1=1.5$, $M_2=-1.63$ at $\tau=0, 2, 25$. 
(a) - wave packet spatial profile, (b) - hodographs, (c) - harmonic spectra, 
(d) - temporal evolution of wave energy, (e) - profiles of the wave 
magnetic fields $b_x$ and $b_y$ at $\tau=25$.}
\label{fig:landsolit}
\end{center}
\end{figure}

It is interesting first to investigate the influence of Landau damping on a
(previously) stable envelope soliton wave train. Figs.\ 
\ref{fig:landsolit}a-\ref{fig:landsolit}e depict
the KNLS waveforms, hodographs, spectra at different times as well as
wave energy degradation and a snapshot of fluctuating fields $b_x$ 
and $b_y$ at $\tau=25$ for a soliton initial profile in $\beta=1$ plasma 
(i.e. $M_1=1.5,~ M_2=-1.63$). 
The effect of the collisionless dissipation is
apparent from the decreasing amplitude and increasing width of the 
waveform (Fig.\ \ref{fig:landsolit}a). Energy ``inverse cascades'' and
accumulates in large scales (low-$k$ harmonics), as shown in Fig.\
\ref{fig:landsolit}c. Note that the temporal energy evolution can
be separated into two phases (see Fig.\ \ref{fig:landsolit}d), namely
a period of rapid damping $0\le\tau\le5$ 
(so that $0\le t\le10^3\Omega_i^{-1}$)
associated with drastic spectral modification, followed by an extended
quasi-stationary period (of duration of, at least, $10^4\Omega_i^{-1}$)
during which the energy of the waveform decays very slowly.

\begin{figure}[p]
\begin{center}
\psfig{file=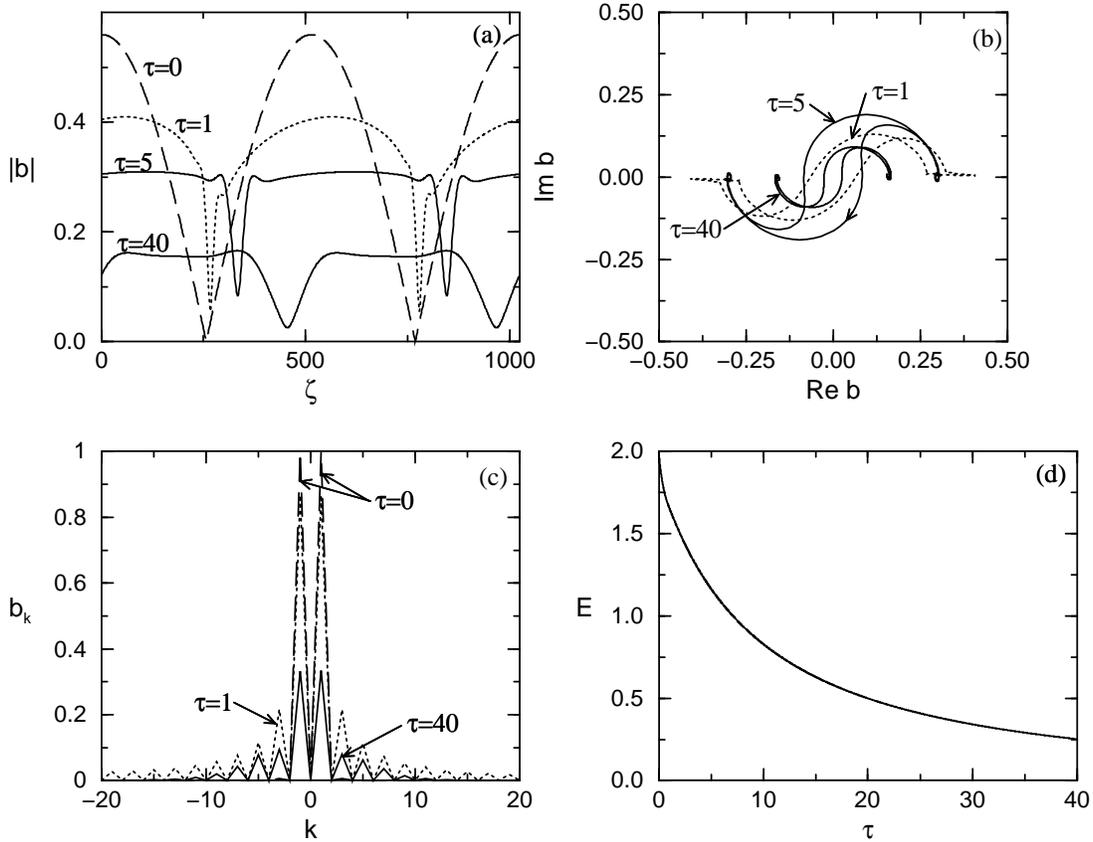,height=9.in}
\vskip-4cm
\caption{Wave evolution from the KNLS with Landau damping for
$\beta=1$, $M_1=1.5$, $M_2=-1.63$ for a parallel propagating,
{\em linearly} polarized sinusoidal wave 
initial condition. (a) - wave profiles, (b) - hodographs, (c) - harmonic
spectra, (d) - temporal evolution of wave energy. At $\tau=0$ only 
$k=1, -1$ harmonics are excited. Note the formation of localized,
 quasi-stationary
waveforms ({\em dissipative structures}) with a characteristic S-shaped
phase diagram and narrow harmonic spectrum.}
\label{fig:landlin}
\end{center}
\end{figure}

The time evolution of an initially linearly polarized parallel propagating wave 
for the same plasma parameters
is shown in Figs.\ \ref{fig:landlin}. Fig.\ \ref{fig:landlin}a depicts
wave amplitude profiles for four times $\tau=0, 1, 5, 40$. 
Localized {\em quasi-stationary} structures are seen to form very rapidly,
the formation time is $\tau_f\sim2$. Again, the amplitude decreases and the
width of the structures increases with time, due to dissipation. Fig.\
\ref{fig:landlin}b represents the phase portrait of the structures at
times $\tau=1, 5, 40$. The dissipative structures exhibit an easily distinguishable
{\em ``S-shaped''} diagram. It is worth to mention that this characteristic 
shape is well preserved for a wide range of plasma parameters and may be
considered a signature of (collisionless) damping (see also Section
\ref{sec:viscous}). The harmonic spectrum of the dissipative structures
is {\em narrow}, with most energy associated with  {\em low-$k$}
harmonics (Fig.\ \ref{fig:landlin}c). For example, for $\tau=40$
the harmonic tail is almost negligible after $|k|\simeq10$.
Energy decay is shown in Fig.\ \ref{fig:landlin}d. No distinct phases 
are seen.

\begin{figure}[p]
\begin{center}
\psfig{file=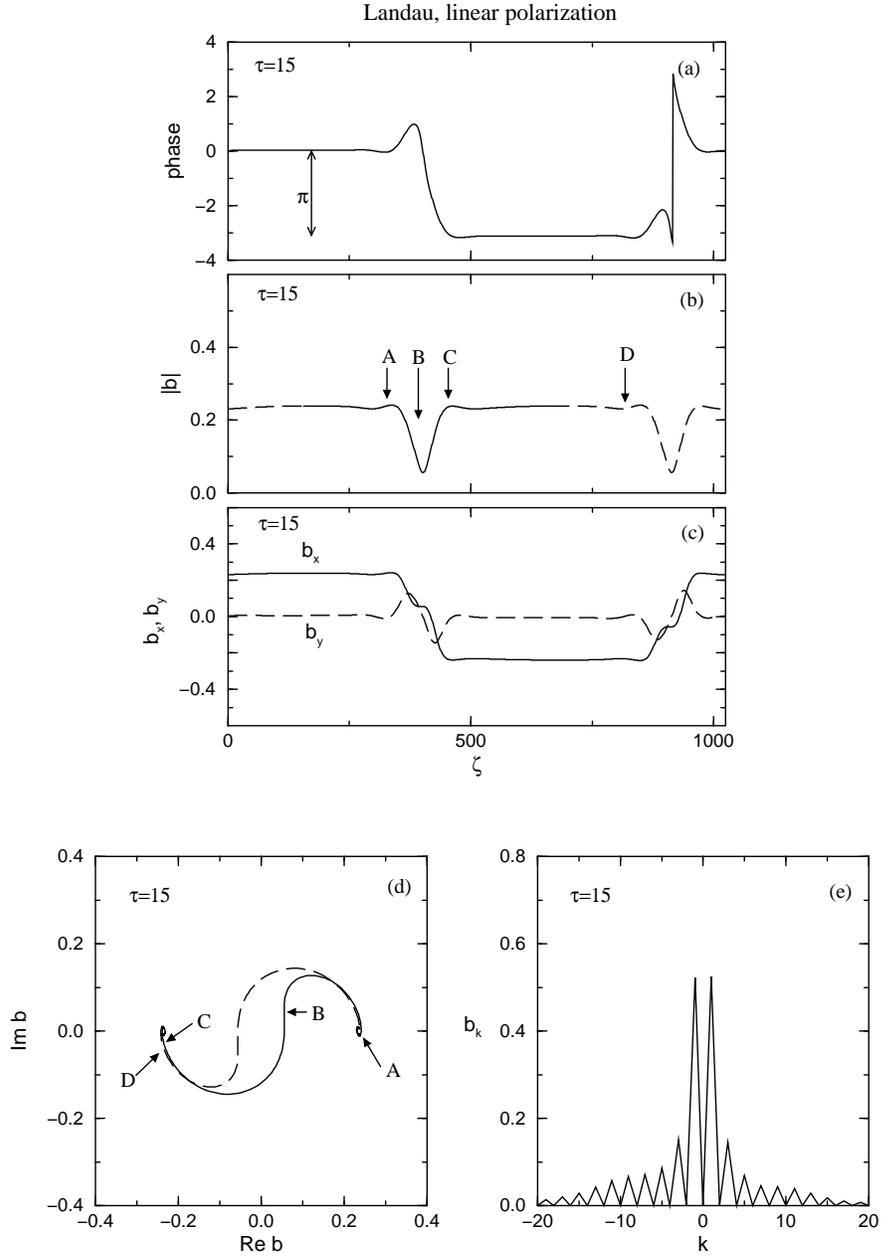,height=9.in}
\vskip-4cm
\caption{Detailed snapshot of a {\em dissipative structure} (or a KNLS
rotational/directional discontinuity) at $\tau=15$ in $\beta=1$ plasma.
(a) - phase profile, (b) - wave amplitude profile, (c) - profiles of the wave
fields $b_x, b_y$ (note narrow spatial localization of the discontinuity), 
(d) - hodograph (note a characteristic {\em S-shape}), (e) -harmonic
spectrum.}
\label{fig:landlin=15}
\end{center}
\end{figure}

\begin{figure}[p]
\begin{center}
\psfig{file=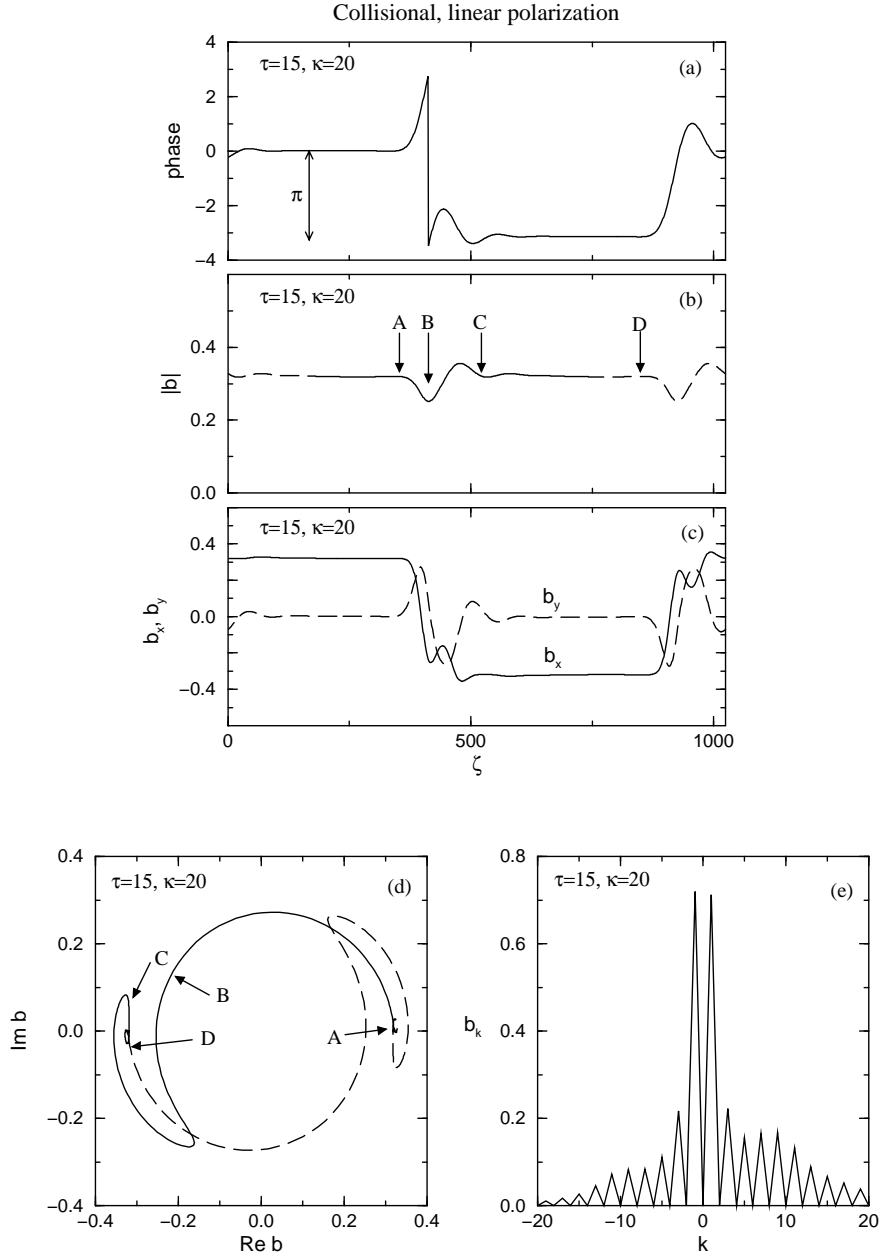,height=9.in}
\vskip-4cm
\caption{Detailed snapshot of a {\em rotational discontinuity} for an
intermediate collisional regime ($\kappa=20$)
at $\tau=15$ in $\beta=0.99$ plasma.
(a) - phase profile, (b) - wave amplitude profile, (c) - profiles of the wave
fields $b_x, b_y$ (note narrow spatial localization of the discontinuity), 
(d) - hodograph (note a typical {\em S-shape}), (e) -harmonic
spectrum.}
\label{fig:visclin+20=15}
\end{center}
\end{figure}

Figs.\ \ref{fig:landlin=15} show the observed localized structures at 
$\tau=15$ in more detail. 
It is seen (Figs.\ \ref{fig:landlin=15}a and
\ref{fig:landlin=15}b) that regions of significant field variations are
accompanied by fast phase rotation through (approximately) $\pi$ radians.
However, in the regions of negligibly varying $|b|$,  linear polarization
is preserved. This is easily seen from comparison of Figs.\
\ref{fig:landlin=15}b and \ref{fig:landlin=15}d. At the discontinuity
(path A-B-C) the magnetic field vector completes a rotation through $\pi$
radians (Fig.\ \ref{fig:landlin=15}d). During the subsequent quescent 
region (path C-D), the magnetic field vector resides at the ``tip'' of  the left
arm indicating pure linear polarization. At the next discontinuity
the vector returns to the initial position, similarly completing a $\pi$
radian rotation, as shown by dashed path in Fig.\ \ref{fig:landlin=15}d. 
Thus, the
KNLS dissipative structures have the requisite properties to be identified 
as localized, {\em rotational/directional discontinuities}, as recently observed 
in the solar wind \cite{DD,S-type}. Note, however, these KNLS 
rotational discontinuities are associated with the regions of {\em varying 
$|b|$}, unlike the conventional definition that $|b|=const$ across the 
rotational
discontinuity. It may be seen from comparison with other cases below
that there is no sharp difference between a rotational and
directional discontinuity. One may be transformed into another by changing
the initial wave polarization and propagation angle. Hence, we use both
words to denote the S-type KNLS discontinuity. We should also note the
remarkable similarity of hodographs obtained by solution of the KNLS
equation and from full numerical plasma simulations \cite{Vasquez1}.
We emphasize the quasi-stationary character of
the KNLS discontinuities, with waveform shape preserved for 
thousands of cyclotron times. The structures exhibit narrow, localized 
spectra (Fig.\ \ref{fig:landlin=15}c).
The inverse spectrum width defines the spatial localization of 
the discontinuity. Also, the small-scale parasitic oscillations
typical of the fluid DNLS are absent (they are probably inhibited
by Landau damping at earlier stages of evolution).
Such KNLS discontinuities occur commonly and are not restricted
to $\beta$'s close to unity. These structures are quite evident in
a wide interval of $\beta$, of approximately 0.5-0.6 to
1.4-1.6. The formation time increases from $\tau_f\sim2$ 
at $\beta\simeq1$ to $\tau_f\sim15$ at the ends of the interval.
Thus, the dissipative structures still form at smaller $M_2$, however
the formation time increases when $M_2$ decreases. Wave profiles
and spectra still look like those in Figs.\ \ref{fig:landlin}, 
\ref{fig:landlin=15}. The phase diagram, however, changes. 
Its central part expands and the ``arms'' become thicker, so that the whole 
diagram become similar to that shown in Fig.\ \ref{fig:visclin+20=15}d
for the case of collisional dissipation. In all cases the phase discontinuity
is localized at the discontinuity of $|b|$.

\begin{figure}[p]
\begin{center}
\psfig{file=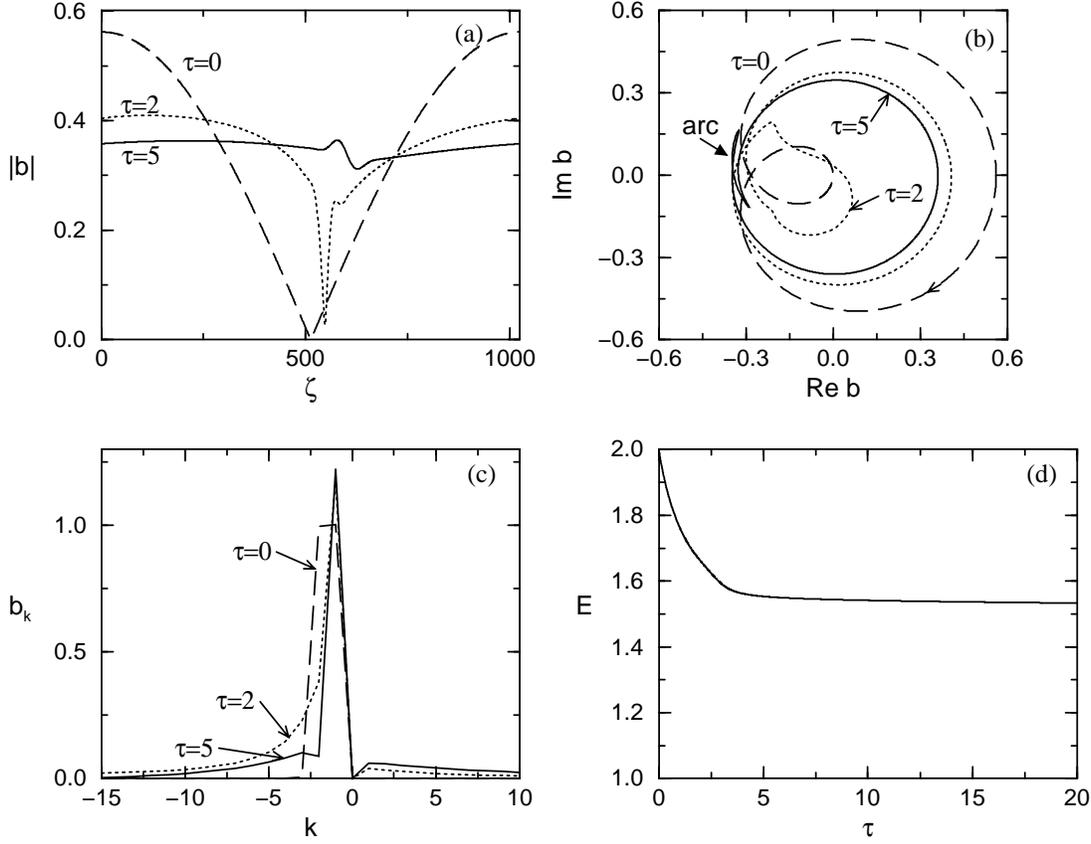,height=9.in}
\vskip-4cm
\caption{Wave evolution for the KNLS with Landau damping for
$\beta=1$, $M_1=1.5$, $M_2=-1.63$ for a  parallel propagating,
{\em circularly} polarized sinusoidal 
wave initial condition. (a) - wave profiles, (b) - hodographs, (c) - harmonic
spectra, (d) - temporal evolution of wave energy. At $\tau=0$ only 
$k=-1, -2$ harmonics are excited. Note the formation of an intermittent,
 arc-polarized discontinuity at $\tau=5$, which disappears quickly due to Landau
damping. Only $k=-1$ circularly polarized harmonic survives and 
experience no steepening or damping.}
\label{fig:landcirc}
\end{center}
\end{figure}

In Figs.\ \ref{fig:landcirc} the corresponding waveform dynamics for an
initially circularly polarized wave with sinusoidal amplitude modulation
is shown for the same plasma parameters. 
In contrast to the previous
case, circularly polarized waves do not evolve into structures with
discontinuities (smooth, wave-like forms emerge). Rather, 
they evolve in a few $\tau$ to a single (almost purely) circularly polarized 
harmonic at the  lowest $k$ (Figs.\ \ref{fig:landcirc}b, \ref{fig:landcirc}c). 
A smooth, almost negligible amplitude modulation is imposed on the wave. 
At these regions, weak, arc-polarized phase irregularities occur (see
Fig.\ \ref{fig:landcirc}b,e). Such phase irregularities are intermittent 
structures living just few $\tau$, they carry little energy (which is mostly 
in higher-$k$ harmonics) and, thus, disappear quickly.
Energy decay (Fig.\ \ref{fig:landcirc}e) is substantial
during the time of spectrum modification, and 
extremely slow at later times.

\begin{figure}[p]
\begin{center}
\psfig{file=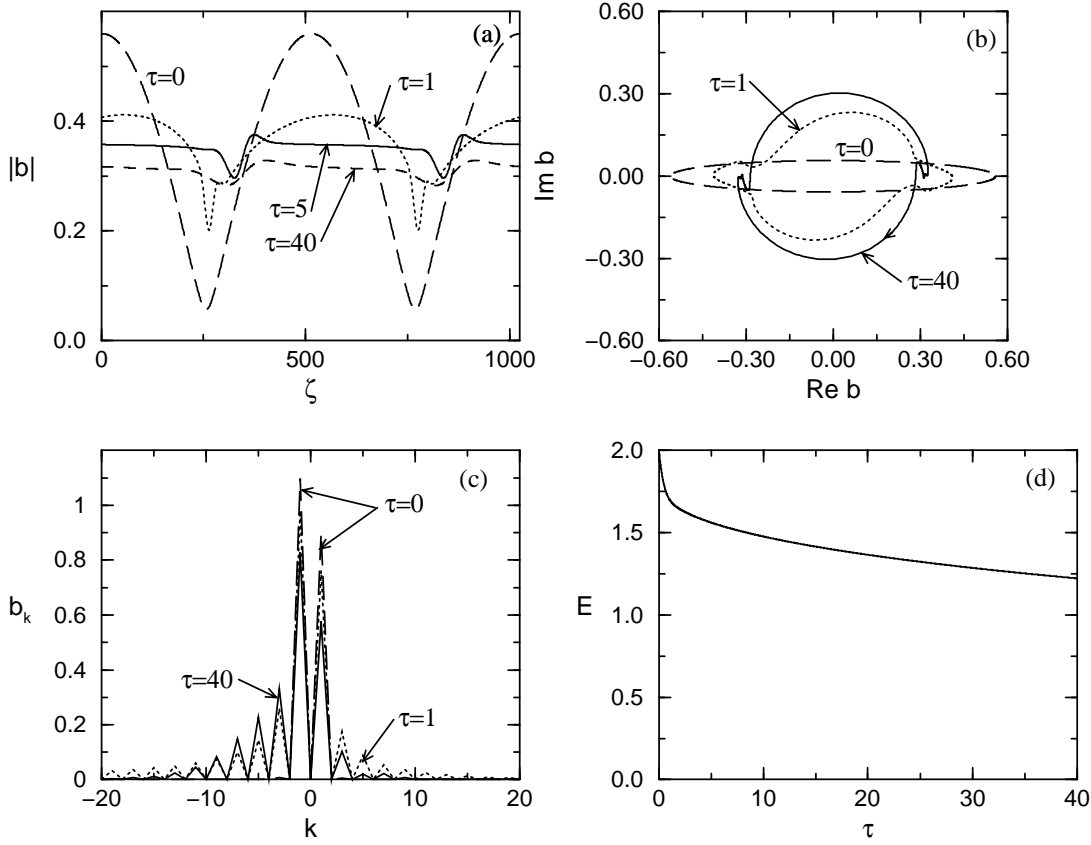,height=9.in}
\vskip-4cm
\caption{Wave evolution for the KNLS with Landau damping for 
$\beta=1$, $M_1=1.5$, $M_2=-1.63$ for a parallel propagating,
{\em elliptically} polarized sinusoidal 
wave initial condition. (a) - wave profiles, (b) - hodographs, (c) - harmonic
spectra, (d) - temporal evolution of wave energy. At $\tau=0$ only 
$k=1, -1$ harmonics are excited and $b_1=0.9, b_{-1}=1.1$. Note the formation
of rotational discontinuities (phase change equal to $\pi$ radians) with 
weak wave amplitude variation across. Note also the smaller energy dissipation
rate than for a linear polarization.}
\label{fig:landell}
\end{center}
\end{figure}

Figs.\ \ref{fig:landell} depict the evolution of an initially elliptically polarized 
wave, an intermediate case between purely circular and linear polarizations.
Plasma parameters are the same as in the previous cases. 
An elliptically polarized wave evolves to
a wave structure of slowly varying amplitude (similar to the circular 
polarization case) as shown in Fig.\ \ref{fig:landell}a. However, there are
sudden phase jumps (by $\pi$ radians) which are localized at regions
of varying wave amplitude (typical of linear polarizations). These phase
discontinuities (which are the semi-circles in Fig.\ \ref{fig:landell}b) are
separated by extended regions of linear polarization during which a
tip of the magnetic field vector resides at small arms on the graph.
Unlike the linear polarization case, the phase discontinuity is not accompanied
by the wave amplitude discontinuity. A harmonic spectrum is again 
sufficiently narrow and most energy is associated with large scales
(see Fig.\ \ref{fig:landell}c). Energy dissipation is weak in comparison to
linear polarizations (Fig.\ \ref{fig:landell}d).

Obliquely propagating waves are still described by the KNLS equation.
However, a new wave field which (formally) contains a perpendicular
projection component of the ambient magnetic field should be introduced.
Assuming the ambient field lies in $x$-$z$-plane, we write the new
field as
\begin{equation}
b=(b_x+B_0\sin{\Theta}+ib_y)/B_0 .
\label{oblique-b}
\end{equation}
The nonlinear evolution of the linearly and highly elliptically polarized waves
is strongly sensitive to the angle between the polarization plane and the plane
defined by the ambient magnetic field vector and the direction of wave
propagation. This angle is set by initial conditions. When this angle is small,
the oscillating wave magnetic field has a longitudinal component
along the ambient field. Thus, we refer such waves to as {\em longitudinal}.
In the opposite case, the wave magnetic field oscillates (nearly)
perpendicularly
to the ambient field. Thus, such waves are called {\em transverse}. Note that
classification fails for circularly polarized waves, since a polarization plane 
cannot be defied in this case.

\begin{figure}[p]
\begin{center}
\psfig{file=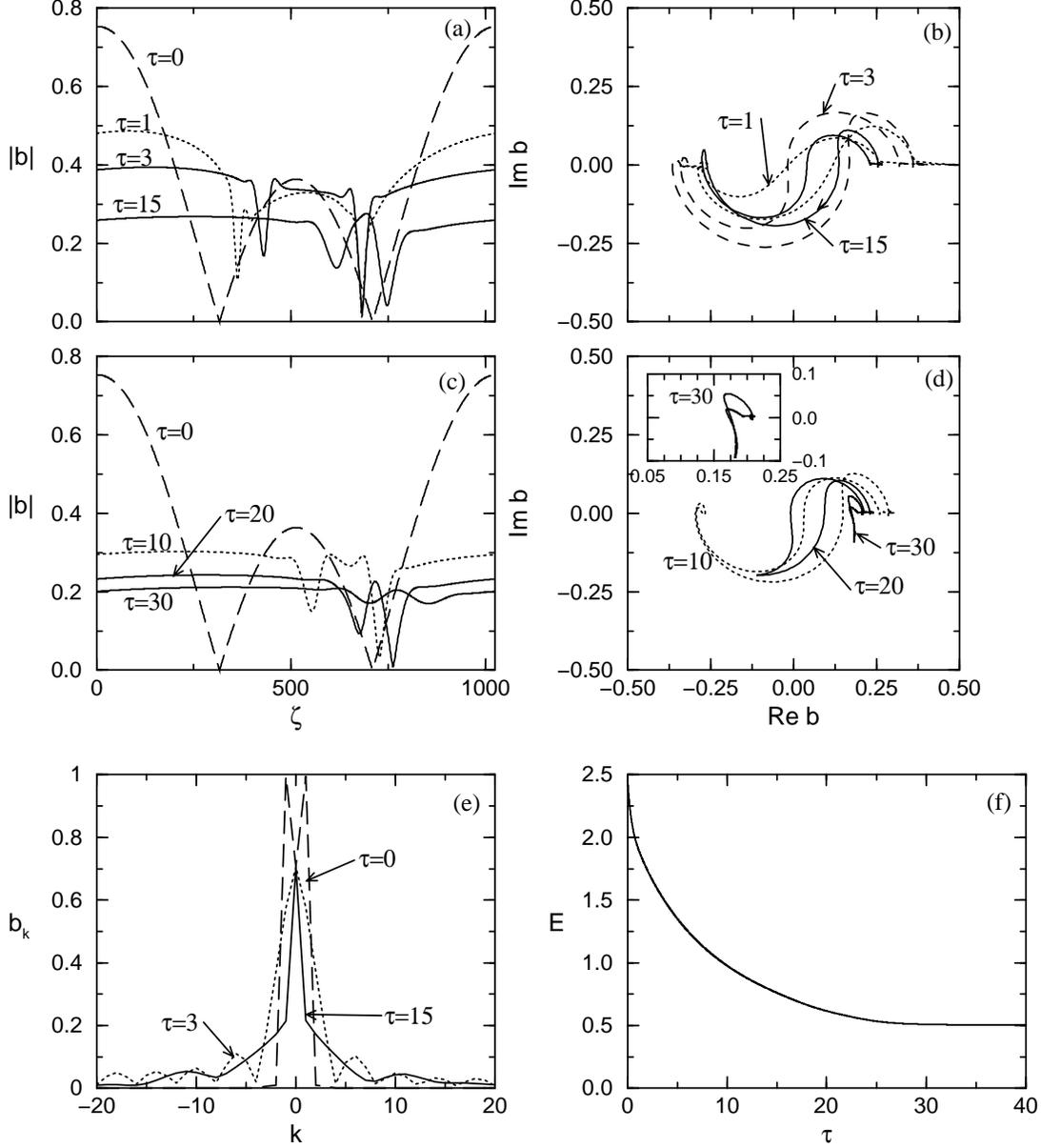,height=9.in}
\vskip-4cm
\caption{Wave evolution for the KNLS with Landau damping for
$\beta=1$, $M_1=1.5$, $M_2=-1.63$ for an {\em obliquely} propagating
($\Theta=45^\circ$), {\em linearly} polarized, {\em longitudinal}, 
sinusoidal wave 
initial condition. (a) and (c) - wave profiles, (b) and (d) - hodographs, 
(e) - harmonic spectra, (f) - temporal evolution of wave energy. 
At $\tau=0$ only $k=1, -1$ harmonics are excited. Note the formation of two 
KNLS directional discontinuities which interact and annihilate
with each other  (at $\tau\simeq20$) to yield a weakly dissipative, 
small amplitude  residual wave structure.}
\label{fig:landofflinlong}
\end{center}
\end{figure}

The time evolution of a transverse wave of initial linear polarization is 
shown in Fig.\
\ref{fig:landofflinlong}, for the same plasma parameters and $\Theta=45^\circ$.
Of course, such a value is near the boundary of applicability of the KNLS.
However, we run such an exaggerated case because 
(i) it takes less computer time, as the nonlinear evolution is faster and, thus, 
(ii) it looks more illustrative than those with smaller $\Theta$.
No significant physics is lost, anyway.
The wave profiles evolve into two KNLS rotational discontinuities
(as in the parallel propagation case) with the typical ``S-polarization''
in a few $\tau$ (see Figs.\ \ref{fig:landofflinlong}a and \ref{fig:landofflinlong}b).
These discontinuities have different group velocities, so some time later
($\tau\sim15$) they start to interact with each other, as seen in Fig.\
\ref{fig:landofflinlong}c. Since these discontinuities are characterized
by opposite phase rotations, they nearly totally annihilate one another 
duting the interaction. The final, residual structure has small amplitude
($b/B_0\simeq0.1$), and is roughly arc-polarized, as seen in Fig.\
\ref{fig:landofflinlong}d at $\tau=30$. This final state is extremely weakly damped
(see Fig.\ \ref{fig:landofflinlong}f) because of the 
very small magnetic field perturbation amplitude of the waveform
(damping in KNLS is nonlinear!). Note here that due to redefinition
of the wave field, Eq.\ (\ref{oblique-b}), the energy formally reads
as $E=E_{wave}+(\sin{\Theta})^2$, where the last term is not associated with 
a wave, but is just a new energy zero-level due to the nonzero perpendicular 
component of the ambient field. The harmonic spectrum evolution is shown in 
Fig.\ \ref{fig:landofflinlong}e.

\begin{figure}[p]
\begin{center}
\psfig{file=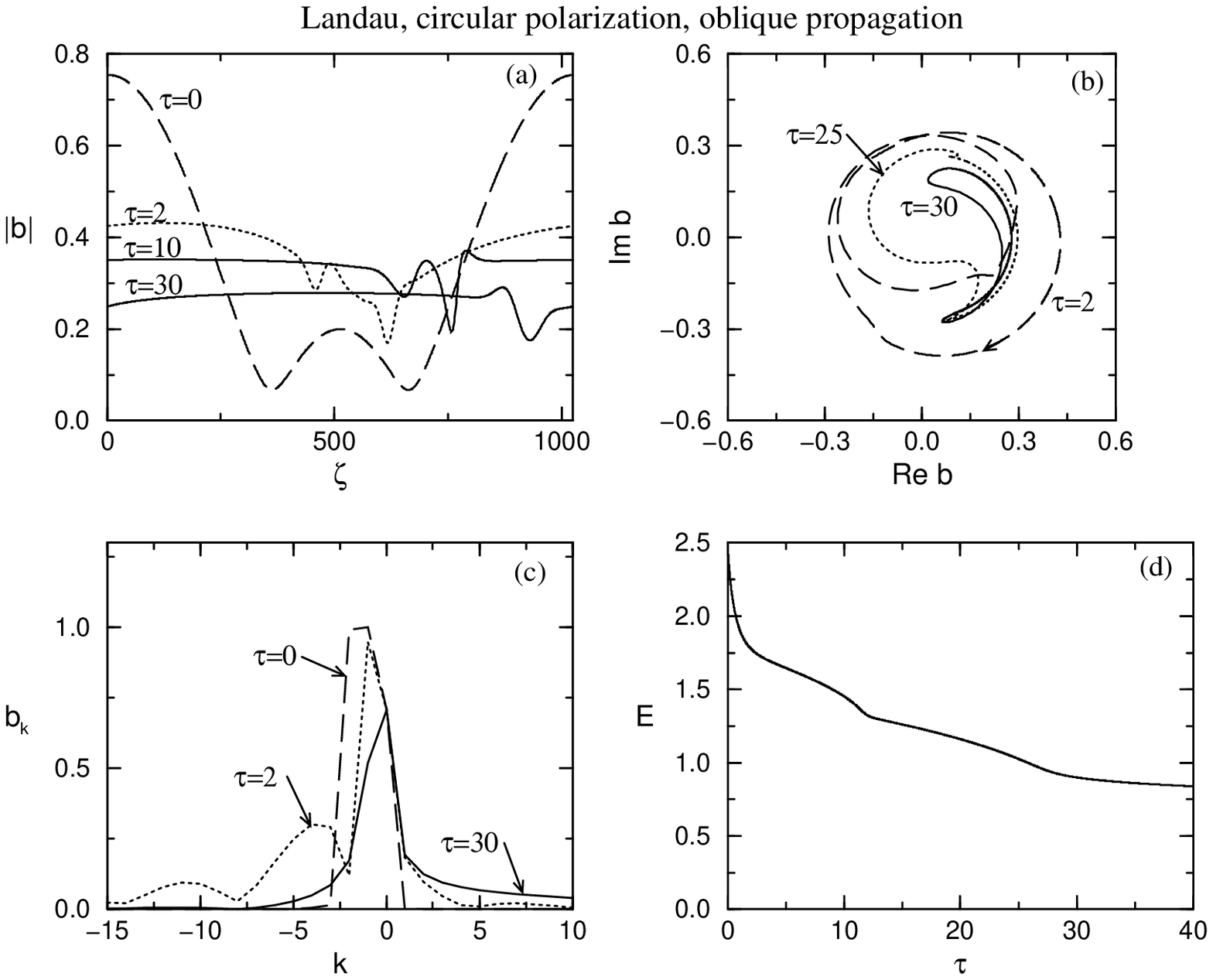,height=9.in}
\vskip-4cm
\caption{Wave evolution for the KNLS with Landau damping for 
$\beta=1$, $M_1=1.5$, $M_2=-1.63$ for an {\em obliquely} propagating,
{\em circularly} polarized, {\em longitudinal}, sinusoidal 
wave initial condition. (a) - wave profiles, (b) - hodographs, (c) - harmonic
spectra, (d) - temporal evolution of wave energy. At $\tau=0$ only 
$k=-1, -2$ harmonics are excited. Note the formation of a stationary 
{\em arc-polarized rotational discontinuity} at approximately $\tau=25$.}
\label{fig:landoffcirc}
\end{center}
\end{figure}

\begin{figure}[p]
\begin{center}
\psfig{file=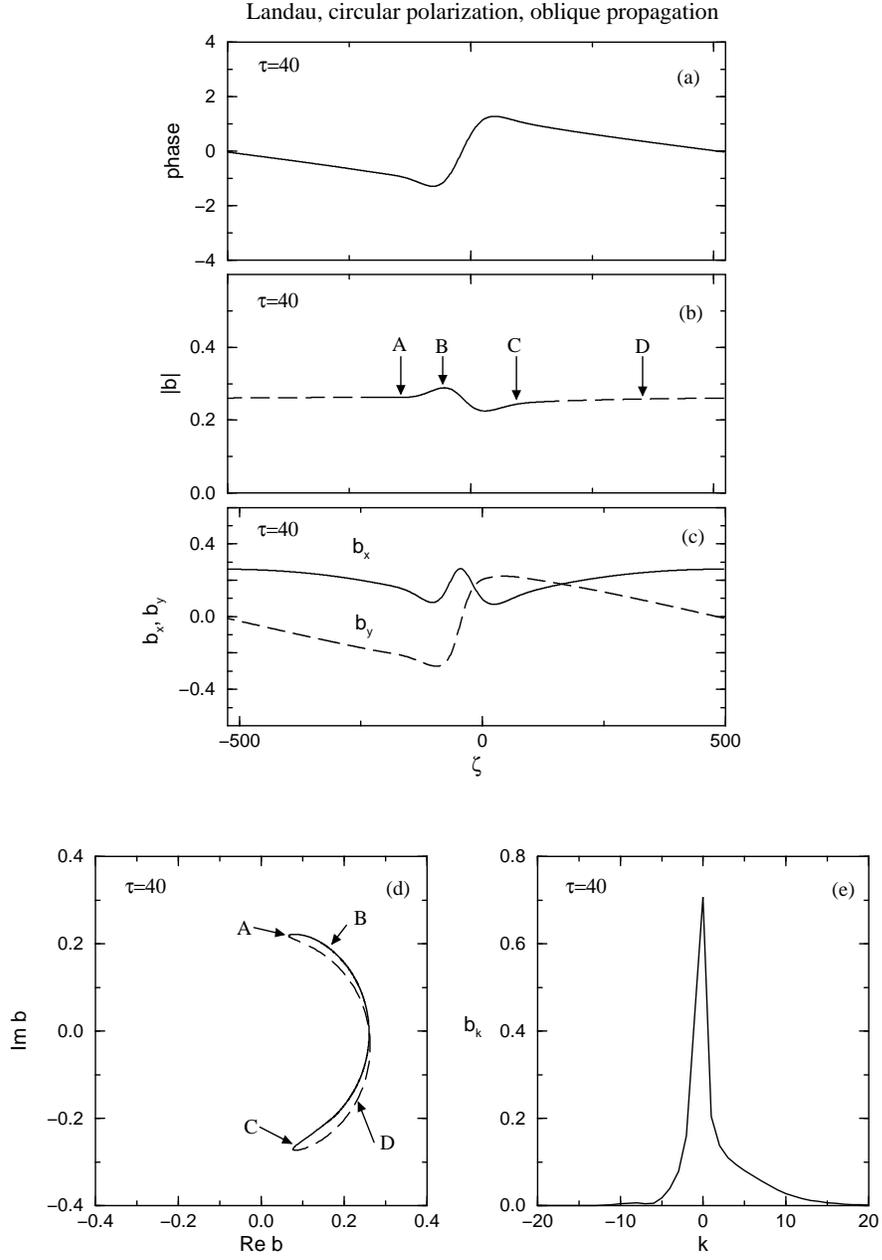,height=9.in}
\vskip-4cm
\caption{Detailed snapshot of an {\em arc-polarized rotational discontinuity} 
at $\tau=40$ for $\beta=1$ plasma.
(a) - phase profile, (b) - wave amplitude profile, (c) - profiles of the wave
fields $b_x, b_y$ (note the narrow spatial localization of the discontinuity
and the negligible wave amplitude modulation),  (d) - hodograph 
(note a characteristic {\em arc-shape}), (e) -harmonic spectrum.}
\label{fig:landoffcirc=40}
\end{center}
\end{figure}

Figs.\ \ref{fig:landoffcirc} show the evolution of an obliquely propagating,
amplitude modulated, circularly polarized wave for the same conditions.
Although some localized structures emerge, they are {\em not} connected
to phase discontinuities. Nevertheless, they have different group velocities
and interact (like in the case of a linear polarization) to yield a weakly
damped, small amplitude, quasi-stationary, purely arc-polarized
waveform (Fig.\ \ref{fig:landoffcirc}b). The amplitude of the waveform is
sufficiently small ($b/B_0\simeq0.2$), but larger than for linear
polarization (note that $|b|/B_0\simeq const$ for such a waveform,
as seen from Fig.\ \ref{fig:landoffcirc}a). Such arc-polarized structures seem
very similar to those recently observed in the solar wind and identified 
with arc-polarized rotational discontinuities \cite{arc}.
We emphasize the almost dissipationless character of such discontinuities
(Fig.\ \ref{fig:landoffcirc}d), inspite of their significant harmonic energy 
content, as seen from the broad spectrum in Figs.\ \ref{fig:landoffcirc}c,
\ref{fig:landoffcirc=40}e.
The energy evolution of oblique circularly polarized waves (unlike all 
other cases) exhibits three different stages (see Fig.\  \ref{fig:landoffcirc}d): 
(i) the primary spectrum modification when two spatially localized wave 
structures emerge ($\tau\le10$), (ii) interaction of these localized
wave structures ($10\le\tau\le30$), and (iii) formation of a residual 
arc-polarized rotational discontinuity ($\tau\ge30$). The different stages 
of spectrum evolution is also shown in Fig.\ \ref{fig:landoffcirc}c.

A (quasi-) stationary, arc-polarized discontinuity is shown in details in Figs.\
\ref{fig:landoffcirc=40}, for $\tau=40$. 
The discontinuity is associated 
with minor (almost negligible) amplitude modulation 
( Fig.\ \ref{fig:landoffcirc=40}b). As seen from Figs.\ 
\ref{fig:landoffcirc=40}a,b,d, the discontinuity is a localized structure of
typically a dispersion width ($ \sim v_A/\Omega_i$). 
The magnetic field vector makes fast clockwise
rotation (see the phase diagram, Fig.\ \ref{fig:landoffcirc=40}d) through
slightly less than $\pi$ radians (path A-B-C). The ends A and C are connected 
by a part of circularly polarized wave packet (slow counterclockwise rotation
in the phase diagram, along the perfect arc C-D-A). Circular polarization is
indicated by smoothly decreasing phase 
outside the discontinuity (Fig.\ \ref{fig:landoffcirc=40}a). 
Since $|b|^2\simeq const$ across the discontinuity
(as well as for a pure circular harmonic), it is nearly decoupled from
dissipation (note, $\widehat{\cal L}[const]=0$, and the second
[dissipative] term in the KNLS vanishes identically). The harmonic spectrum 
of the discontinuity depicted in Fig.\ \ref{fig:landoffcirc=40}e is quite 
narrow and of low amplitude (remember, the $b_0$ harmonic corresponds 
to the ambient field, not the wave). Note the  remarkable similarity of this
solution of the KNLS equation to the structures detected in the solar wind
and observed in computer simulations \cite{arc,Vasquez3}.

The time evolution of an elliptically polarized, obliquely propagating, 
longitudinal
wave (not shown) is very similar to the case of parallel propagation until
large times (i.e. $\tau\sim60$). At later times ($\tau>60$), an arc-polarized
rotational discontinuity forms. There the magnetic field vector rotates rapidly
rotation through an angle of less than $\pi$ radians.

\begin{figure}[p]
\begin{center}
\psfig{file=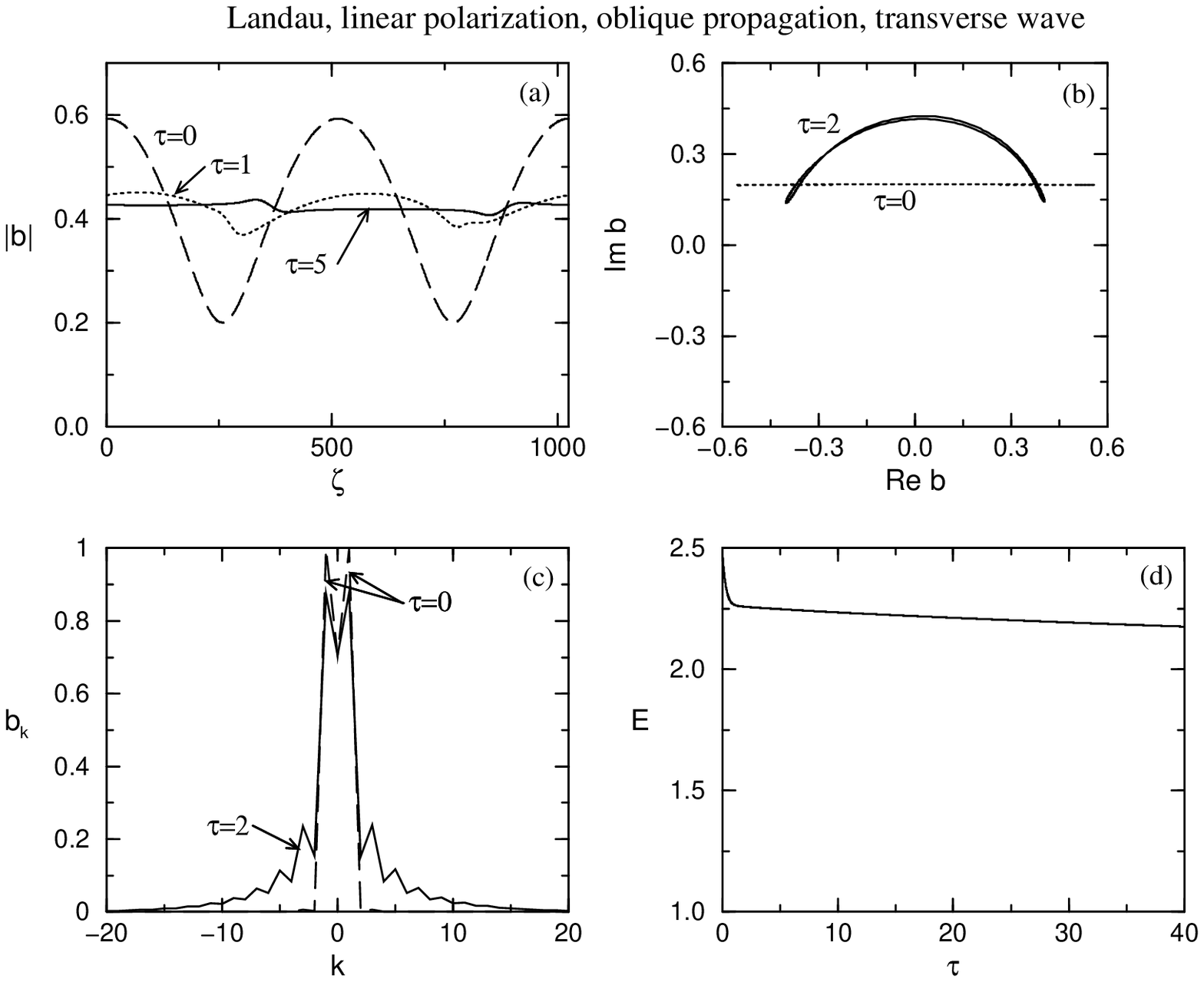,height=9.in}
\vskip-4cm
\caption{Wave evolution for the KNLS with Landau damping for 
$\beta=1$, $M_1=1.5$, $M_2=-1.63$ for an {\em obliquely} propagating,
{\em trasversal}, sinusoidal 
wave initial condition. (a) - wave profiles, (b) - hodographs, (c) - harmonic
spectra, (d) - temporal evolution of wave energy. At $\tau=0$ only 
$k=1, -1$ harmonics are excited. Note the formation of a stationary 
{\em arc-polarized rotational discontinuity} at approximately $\tau=2$.}
\label{fig:landofflintrans}
\end{center}
\end{figure}

The nonliear evolution of the transverse waves is complitely differennt form
that of the longitudinal waves, as shown in Figs.\ \ref{fig:landofflintrans}. 
Such waves quickly evolve (i a few $\tau$)
to an arc-polarized rotational discontinuity (Fig.\ \ref{fig:landofflintrans}a,b).
The magnetic field vector rotates
through an angle less than $180^\circ$ at the discontinuity.
Energy dissipation is negligible in this case (Fig.\ \ref{fig:landofflintrans}d). 
Harmonic spectra of these wveforms are broad  (Fig.\ \ref{fig:landofflintrans}c)

\subsection{The two mode coupling model}

To  {\em qualitatively} understand the physical processes which underly the 
behavior of dissipative nonlinear Alfv\'en waves (of linear and elliptical
polarization), we construct a simple two-mode coupling model.
We assume the spectrum consists of three low-$k$ harmonics
$k=0, \pm1$ (a finite amplitude $k=0$ harmonic corresponds to
oblique propagation, $b_0=B_\bot/B_0=\sin{\Theta}$. This model is based on
the results of numerical solutions which show that only two (initially
excited) $k=\pm1$ harmonics dominate the spectra. Other higher-$k$
spectrum components are of relatively small amplitude and thus can
be neglected. We also neglect linear dispersion which is significant
at the steep fronts, i.e. for broad spectra, only. Here we emphasize once again
that we do not pretend to give a quantitatively correct answer, but only 
intend to get some insight into the essential physics.

We write the wave magnetic field in the form $b=b_+e^{i\zeta}+b_0
+b_-e^{-i\zeta}$ and substitute it into the KNLS equation, Eq.\ 
(\ref{knls-num}), together with Eq.\ (\ref{u-landau}). The system of 
evolution equations for each harmonic $b_j$ is
\begin{mathletters}
\begin{eqnarray}
& &\dot b_++2i(M_1+iM_2)\left[|b_0|^2(b_++b_-^*)+b_+|b_-|^2\right]=0 , 
\label{b+} \\
& &\dot b_0=0 , 
\label{b0} \\
& &\dot b_--2i(M_1-iM_2)\left[|b_0|^2(b_+^*+b_-)+|b_+|^2b_-\right]=0 .
\label{b-}
\end{eqnarray}
\end{mathletters}
We first consider the case of parallelly propagating waves, so that 
$b_0=\sin{\Theta}=0$. The solution of the second equation is simply
$b_0=const$. Dividing the  first equation by the third, we obtain the useful 
equation
\begin{equation}
\frac{{\rm d}b_+}{{\rm d}b_-}
=-\frac{(M_1+iM_2)}{(M_1-iM_2)}\frac{b^*_-}{b^*_+} .
\label{d-conserv}
\end{equation}
While a general solution can be found, we separately consider (for simplicity) the
cases $M_1=0$ and $M_2=0$.

\subsubsection{No dissipation, $M_2=0$}

In the dissipationless (fluid DNLS) limit (i.e. $\beta\to0$), we have from
Eq.\ (\ref{d-conserv}) 
\begin{equation}
|b_+(\tau)|^2+|b_-(\tau)|^2\equiv E_*=const
\label{conservE}
\end{equation}
which is just  conservation of energy. The energy $E_*$ is defined by
an initial condition. Using Eq.\ (\ref{conservE}), the system of Eqs.\
(\ref{b+})-(\ref{b-}) can be easily integrated to yield
\begin{equation}
b_\pm(\tau)=b_\pm(0)~e^{\mp i\omega_\pm\tau} ,
\end{equation}
where $\omega_\pm=2M_1|b_\mp(0)|^2\sim M_1 E_*$ .
This solution represents resonant energy exchange between the two
harmonics. The typical time scale is of order of the shock formation 
time and is approximately equal to 
\begin{equation}
\tau_s\sim (M_1 E_*)^{-1} .
\label{t-s}
\end{equation}

\subsubsection{Strong damping, $M_1\ll M_2$}

In the case of strong damping (i.e. $\beta\simeq1$), we neglect by 
the term $M_1$. Then Eq.\ (\ref{d-conserv}) yields
\begin{equation}
|b_+(\tau)|^2-|b_-(\tau)|^2\equiv \Delta=const .
\eqnum{\ref{conservE}$'$}
\label{conservD}
\end{equation}
Unlike the previous case, Landau damping conserves the energy
difference associated with different Fourier harmonics, so that the
$+k\to-k$ symmetry of the spectrum is preserved. Integration
of Eqs.\ (\ref{b+})-(\ref{b-}) with Eq.\ (\ref{conservD})
results in ($\Delta\not=0$)
\begin{mathletters}
\begin{equation}
|b_+(\tau)|^2=|b_+(0)|^2{\Delta\over \left(1+\Delta/|b_+(0)|^2\right)
\exp{\left(8|M_2\Delta|\tau\right)}-1}
\end{equation}
and ($\Delta\to 0$)
\begin{equation}
|b_+(\tau)|^2=\frac{|b_+(0)|^2}{1+8|M_2|\tau} .
\end{equation}
\end{mathletters}
Initially elliptical polarizations correspond to $b_+\not=b_-$, thus
$$
|\Delta|=\Bigl||b_+(0)|^2-|b_-(0)|^2\Bigr|>0 .
$$
 Thus, harmonic
amplitudes decay exponentially at larger times, 
\begin{equation}
b_\pm\propto e^{-4|M_2\Delta|\tau}
\label{b-ell}
\end{equation}
on the characteristic decay time scale of
\begin{equation}
\tau_{decay}^{ell}\sim|M_2\Delta|^{-1}
\label{t-ell}
\end{equation}
By contrast, initially linear polarizations with $b_+\simeq b_-$
(i.e. $\Delta\to 0$) decay as a square root of time,
\begin{equation}
b_\pm\propto 1/\sqrt{|M_2|\tau}
\eqnum{\ref{b-ell}$'$}
\end{equation}
with the faster characteristic time-scale
\begin{equation}
\tau_{decay}^{lin}\sim|M_2|^{-1}
\eqnum{\ref{t-ell}$'$}
\end{equation}

\subsubsection{Quasi-perpendicular propagation}

Now we neglect cubic terms $b_\pm|b_\mp|^2$ in Eqs.\
(\ref{b+})-(\ref{b-}) as small compared to $b_\pm|b_0|^2\simeq b_\pm$.
Then these equations read
\begin{mathletters}
\begin{eqnarray}
& &\dot b_++2i(M_1+iM_2)|b_0|^2(b_++b_-^*)=0 , 
\\
& &\dot b_--2i(M_1-iM_2)|b_0|^2(b_+^*+b_-)=0 .
\end{eqnarray}
\end{mathletters}
These yield:
\begin{equation}
\frac{\rm d}{{\rm d}\tau}(b_++b_-^*)+4i(M_1+iM_2)(b_++b_-^*)|b_0|^2=0 .
\end{equation}
It is easily seen that harmonic amplitudes decay exponentially
\begin{equation}
b_\pm\propto e^{-4|M_2|\sin^2\!{\Theta} \tau}
\end{equation}
and the characteristic decay time is
\begin{equation}
\tau_{decay}^{obl}\sim |M_2\sin^2\!{\Theta}|^{-1}\le
\tau_{decay}^{lin}\ll\tau_{decay}^{ell}
\end{equation}
Thus, given an isotropic initial distribution, wave excitation
in the quasi-parallel direction is most robust while waves in
the quasi-perpendicular 
direction are quickly damped.

\begin{figure}[p]
\begin{center}
\psfig{file=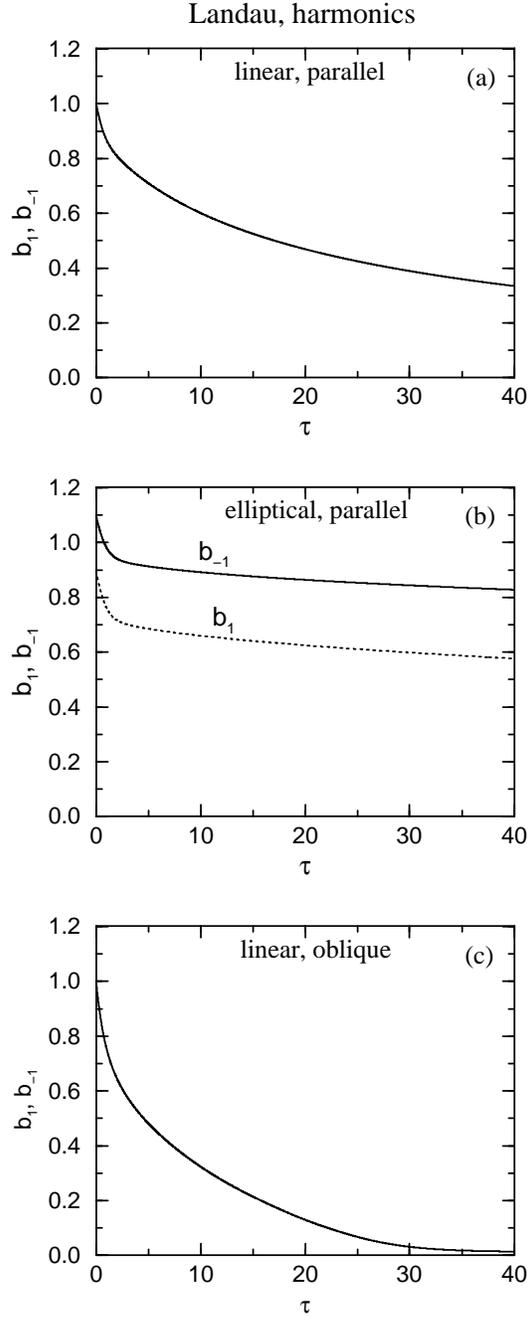,height=9.in}
\vskip-4cm
\caption{Temporal evolution of two initially excited harmonics for the
KNLS for $\beta=1$ for a (a) parallel propagating, linearly polarized wave
($b_{-1}$ and $b_1$ harmonics overlaid), (b) parallel propagating,
elliptically polarized wave, and (c) obliquely propagating 
($\Theta=45^\circ$), linearly polarized wave.}
\label{fig:harmonics}
\end{center}
\end{figure}

Figs.\ \ref{fig:harmonics} represent the time evolution of two $k=\pm1$
harmonics for the cases shown in Figs.\ \ref{fig:landlin}, \ref{fig:landell},
and \ref{fig:landoffcirc}, respectively.
Fig.\ \ref{fig:harmonics}a corresponds to linear polarization (i.e. $\Delta=0$). 
The harmonics coincide all the time and decay according to $\sim1/\sqrt{\tau}$.
Fig.\ \ref{fig:harmonics}b is the case of elliptical polarization
($|\Delta|=0.4$). As seen from this figure, $\Delta$ is approximately
constant and the harmonics decay exponentially, contrary to the case of linear
polarization. Fig.\ \ref{fig:harmonics}c shows the case of obliquely
($\Theta=45^\circ$) propagating, circularly polarized waves.

Amplitude modulated, parallel propagating circularly polarized waves 
with initially excited harmonics $-1$ and $-2$ can be considered
analogously. However, one should consider a spectrum which includes  $-2\le k\le+2$
components.  This complicates the analytical treatment of the problem.
The characteristic time, however, can be estimated from the
elliptical polarization case, Eq.\ (\ref{t-ell}). Now $\Delta\simeq |b_{-2}|^2$,
and the second (i.e. $b_{-2}$) harmonic  dissipates in a time
\begin{equation}
\tau_{decay}^{circ}\sim \left(|M_2 ||b_{-2}|^2\right)^{-1} .
\end{equation}
After this time, only one $b_-$ harmonic survives. Thus, no significant 
damping occures.

\subsection{Discussion}

The sharp contrast between the nonlinear evolution of initially linearly
(elliptically) and circularly polarized parallel propagating wave trains is an
immediate consequence of the unique harmonic scaling of collisionless
(Landau) dissipation in the KNLS equation. Integration of Eq.\
(\ref{mainKNLS}) with $U_{NL}$ defined by Eq.\ (\ref{u-landau}) over all
space (assuming that $b$ vanishes at infinity) yields (in dimensionless
variables)
\begin{eqnarray}
{\partial E\over\partial\tau}&=&-M_2\int_{-\infty}^\infty
|b|^2\frac{\partial}{\partial\zeta}\widehat{\cal L}\left[|b|^2\right] 
{\rm d}\zeta
\nonumber\\
&=&-M_2\int_{-\infty}^\infty 
\left(\frac{1}{2\pi}\int_{-\infty}^\infty e^{ik'\zeta}
\left(|b|^2\right)_{k'} {\rm d}k'\right) \nonumber\\
& &{ }\cdot\frac{\partial}{\partial\zeta}
\left(\int_{-\infty}^\infty e^{ik''\zeta}~i\frac{k''}{|k''|}
\left(|b|^2\right)_{k''} {\rm d}k''\right) {\rm d}\zeta
\nonumber\\
&=&\frac{M_2}{2}\int_{-\infty}^\infty 
\left|\left(|b|^2\right)_k\right|^2|k|~ {\rm d}k~ ,
\label{dE-landau}
\end{eqnarray}
where $E=\int_{-\infty}^\infty |b|^2 {\rm d}\zeta$ is the total wave energy,
and we assumed $\langle|b|^2\rangle=0$.(This result have been 
reported also in Refs.\ \cite{MW,WFlaM}.)
Thus, the rate of energy dissipation is controlled by the coupling coefficient,
$M_2<0$. Also, we conclude that higher-$k$ harmonics are strongly
damped which is typical for a phase-mixing process (i.e. smaller
scales mix faster). It is crucial, however, to understand that 
{\em collisionless damping enters at all $k$}, in contrast to hydrodynamic
systems where diffusion (viscosity) yields dissipation only at large $k$
(i.e. small scales, or steep gradient regions). Hence, competition of nonlinear 
harmonic generation (i.e. front steepening) and collisionless damping
produces a quasi-stationary energy spectrum which is a {\em rapidly}
decreasing function of $k$, with the long-time asymptotic behavior
dominated by a few low-$k$ harmonics. Since linear polarizations have
(initial) spectra symmetric upon $k\to-k$, they couple more strongly
to dissipation than circular polarizations do. Moreover, interaction of finite
amplitude spectrum components necessarily leads to higher-$k$
harmonic generation and wave steepening. Such interactions are excluded
for pure circular polarization. Therefore, the circularly polarized wave
evolves to a single harmonic final state, which is, itself, a stationary 
(and exact) solution of the KNLS equation (i.e. it experiences no steepening
and minimal damping).

Since Landau damping enters symmetrically for $+k$ and $-k$ spectrum
components, it does not change the initial symmetry of a spectrum,
so that the sense of an initial polarization (set by initial spectrum symmetry)
 is preserved (neglecting feeble high-$k$, low-amplitude asymmetry induced 
by steepening). Thus, KNLS evolution preserves initial wave
helicities forming directional/rotational discontinuities at the edges 
of the regions of linear (elliptical) polarization. These 
directional/rotational discontinuities are localized at regions 
of strong amplitude
variations, unlike the conventional view \cite{Cohen} where the shock and
rotational discontinuity are well separated and have different group velocities.
The KNLS directional/rotational discontinuities display a well distinguished 
S-shaped phase portrait which can be considered a footprint of the occurrence of 
strong Landau damping in satellite observations. Since the KNLS equation 
preserves (initial) wave helicity \cite{Dawson,FlaMW}
, rotational discontinuities do not form from 
circularly polarized waves. Given that circular polarizations 
decay much more slowly than linear polarizations do, the results
suggest that large-scale circularly polarized waves should occur more
frequently than small scale discontinuities which evolve from linear
polarizations (given comparable initial populations), especially at large
distances from the Sun.

Oblique and quasi-perpendicular, linearly polarized waves first evolve to two 
KNLS directional/rotational discontinuities of opposite phase rotation 
propagating with
different group velocities, so that they finally merge and annihilate to yield
weakly-nonlinear, small amplitude, arc-polarized waves. Despite the fact that 
oblique, circularly polarized waves do not form KNLS rotational 
discontinuities, their final state is similar to (but of larger amplitude than)
that of a linear polarization.
Since, oblique and quasi-perpendicular (both linear and circular 
polarizations) waves are coupled to the ambient magnetic field 
$B_\bot=B_0\sin{\Theta}$, which is usually (much) larger than the 
fluctuating wave field $b$, the process of energy dissipation is
$B_\bot^2/b^2$ times faster for such waves. Waveforms again evolve 
to the lowest-$k$ state which is now the single harmonic 
$b_0=B_\bot/B_0$ (which corresponds to the ambient field component,
not the wave perturbed field) indicating dissipation of {\em all wave energy}.
Thus, the final state of nonlinear evolution for ``off-axis'' propagating waves
is a (very) small amplitude, weakly-nonlinear, {\em weakly-damped}, 
arc-polarized Alfv\'en wave packet. Pure arc-polarized waveforms (which 
develop from circular polarizations) have much in common with the
structures recently observed in the solar wind and identified with 
arc-polarized rotational discontinuities \cite{arc}. Such structures 
experience minimal damping. 
The comparably fast damping of oblique and
quasi-perpendicular waves suggests that most magnetic fluctuations
should occur at small angles to the ambient field (given initial isotropic 
fluctuations) resulting in large turbulence levels in parallel and 
quasi-parallel directions and small turbulence levels in quasi-perpendicular 
directions where Alfv\'enic turbulence is sufficiently suppressed.

\section{Collisional dissipative Alfv\'en trains}
\label{sec:viscous}
\subsection{Numerical solutions}

The evolution equation for coherent Alfv\'en waves is again Eq.\ (\ref{mainKNLS})
with the nonlinear velocity perturbation term written as
\begin{equation}
U_{NL}=\frac{v_A}{2}\left\{Q_1\left(|b|^2-\langle|b|^2\rangle\right)
+Q_2{\widehat{\cal J}}_L\left[|b|^2-\langle|b|^2\rangle\right]\right\} ,
\label{u-viscous}
\end{equation}
where
\begin{eqnarray*}
Q_1&=&{1\over (1-\beta/\gamma)} , \\
Q_2&=&-\frac{\beta}{\chi^c_\|} {\gamma-1\over\gamma}
{1\over (1-\beta/\gamma)^2 } ,
\end{eqnarray*}
$\gamma=3$, and $\chi^c_\|$ is the parallel heat conduction coefficient 
which represents diffusive (collisional) energy dissipation. Here 
${\widehat{\cal J}}_L$ is the collisional dissipation (integral) operator
\begin{equation}
{\widehat{\cal J}}_L[f](x)=e^{-x/L}\int^x e^{x'/L} f(x'){\rm d}x'
\doteq\frac{1}{1/L-ik} ,
\end{equation}
where $L=(\chi^c_\|/v_A)(1-\beta/\gamma)/(1-\beta)$. For future use
we introduce $\kappa=\bar z/L$, the dimensionless dissipation scale, i.e.
the ratio of a typical wave length to the dissipation length. For $\beta$ close
to unity we may write approximately $\kappa\simeq [75(1-\beta)/\chi^c_\|]
(v_A^2/\Omega_i)$.

\begin{figure}[p]
\begin{center}
\psfig{file=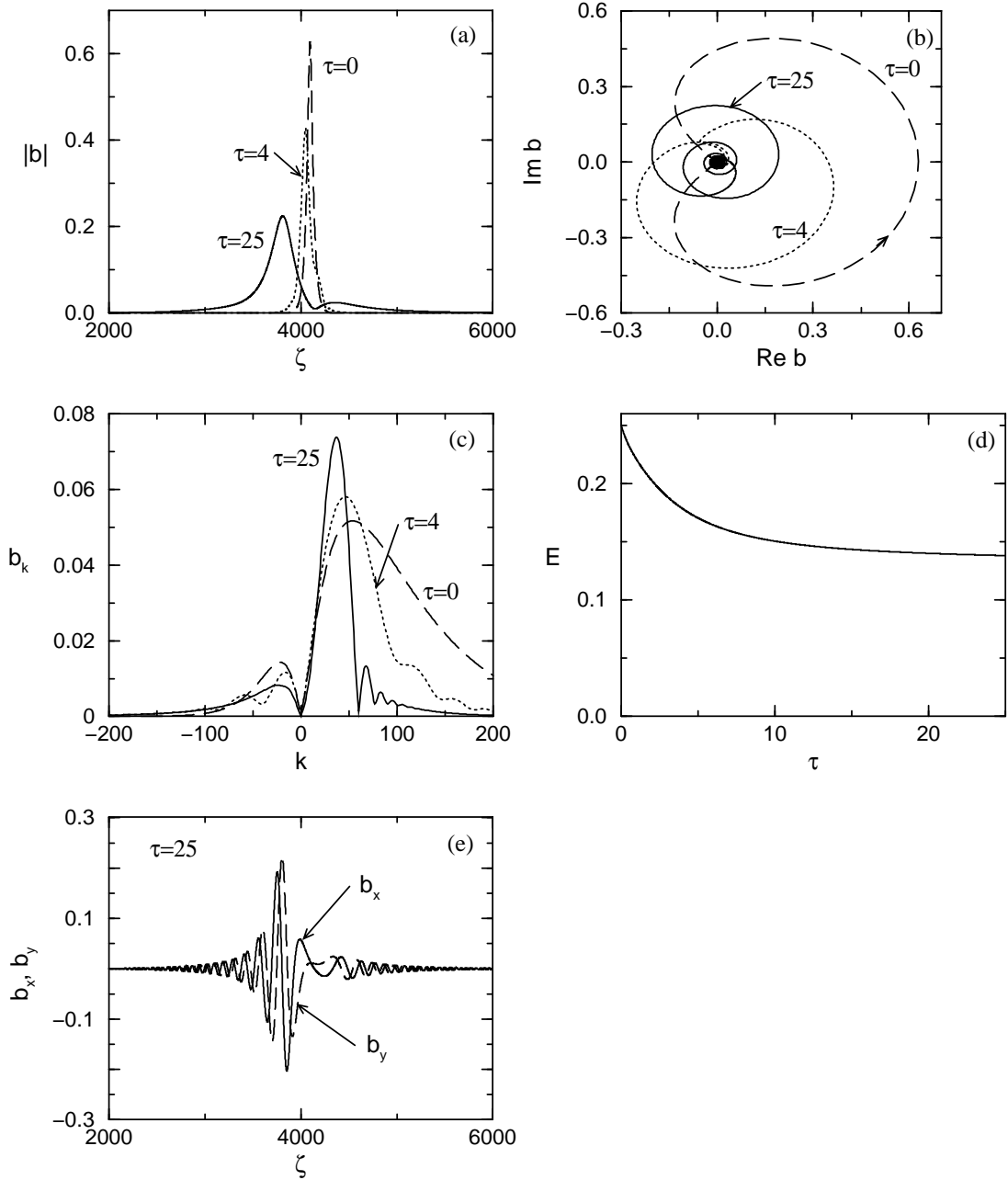,height=9.in}
\vskip-4cm
\caption{Evolution for a parallel propagating, 
{\em soliton} initial condition for the KNLS with 
collisional damping for $\beta=0.99$ and $\kappa=100$ 
($\kappa\simeq k_\textrm{typical}$
corresponds to the strongest damping) at $\tau=0, 4, 25$. 
(a) - wave packet spatial profile, (b) - hodographs, (c) - harmonic spectra, 
(d) - temporal evolution of wave energy, (e) - profiles of the wave 
magnetic fields $b_x$ and $b_y$ at $\tau=25$.}
\label{fig:viscsolit+100}
\end{center}
\end{figure}

As in the case of collisionless dissipation, 
we first investigate the temporal evolution 
of an (initially) soliton wave envelope, Eqs.\ (\ref{solit-a})-(\ref{solit-b}),
since it is stable in absence of any dissipation. Solutions at different times are 
shown in Figs.\ \ref{fig:viscsolit+100} with $\kappa=100$ and $\beta=0.99$.
We shall see that collisional dissipation is strongest when $\kappa\sim k$,
i.e. when the dissipation scale is comparable with a typical wave-length 
of a train. We again (as with Landau damping) see rapid primary
spectrum and spatial profile modification (Figs.\ \ref{fig:viscsolit+100}c,a,
respectively) and formation of a quasi-stationary solution with negligible
energy degradation (Fig.\ \ref{fig:viscsolit+100}d). The hodographs are shown 
at three times $\tau=0,4,25$ in Fig.\ \ref{fig:viscsolit+100}b. The steady-state
profiles of $b_x$ and $b_y$ are shown separately in Fig.\ \ref{fig:viscsolit+100}e.
We should mention that the effect of collisional damping is significant only
for $\beta$ very close to unity.

\begin{figure}[p]
\begin{center}
\psfig{file=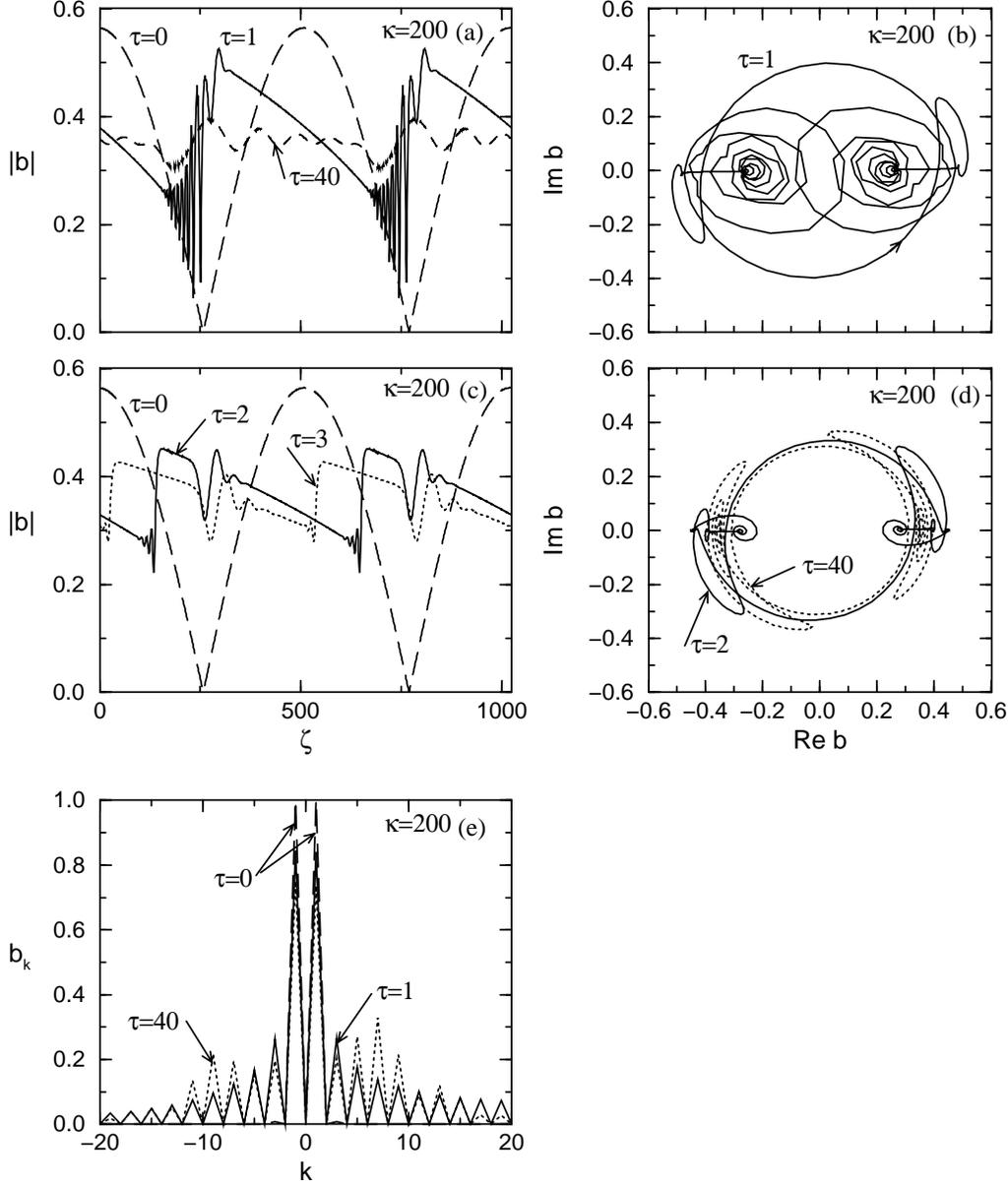,height=9.in}
\vskip-4cm
\caption{Wave evolution for the KNLS with collisional damping with 
$\beta=0.99$ and $\kappa=200$ (weakly collisional regime) for parallel 
propagating, {\em linearly} polarized sinusoidal wave 
initial condition. (a) and (c) - wave profiles, (b) and (d) - hodographs, 
(e) - harmonic spectra. At $\tau=0$ only 
$k=1, -1$ harmonics are excited. Note the formation of propagating shock waves
(sharp fronts) and rotational discontinuities (wider wave-like forms).
The noisy parasitic oscillations generated at earlier times ($\tau=1$) are possibly 
due to modulation instability, which is stronger for smaller scale modulations.}
\label{fig:visclin+200}
\end{center}
\end{figure}

\begin{figure}[p]
\begin{center}
\psfig{file=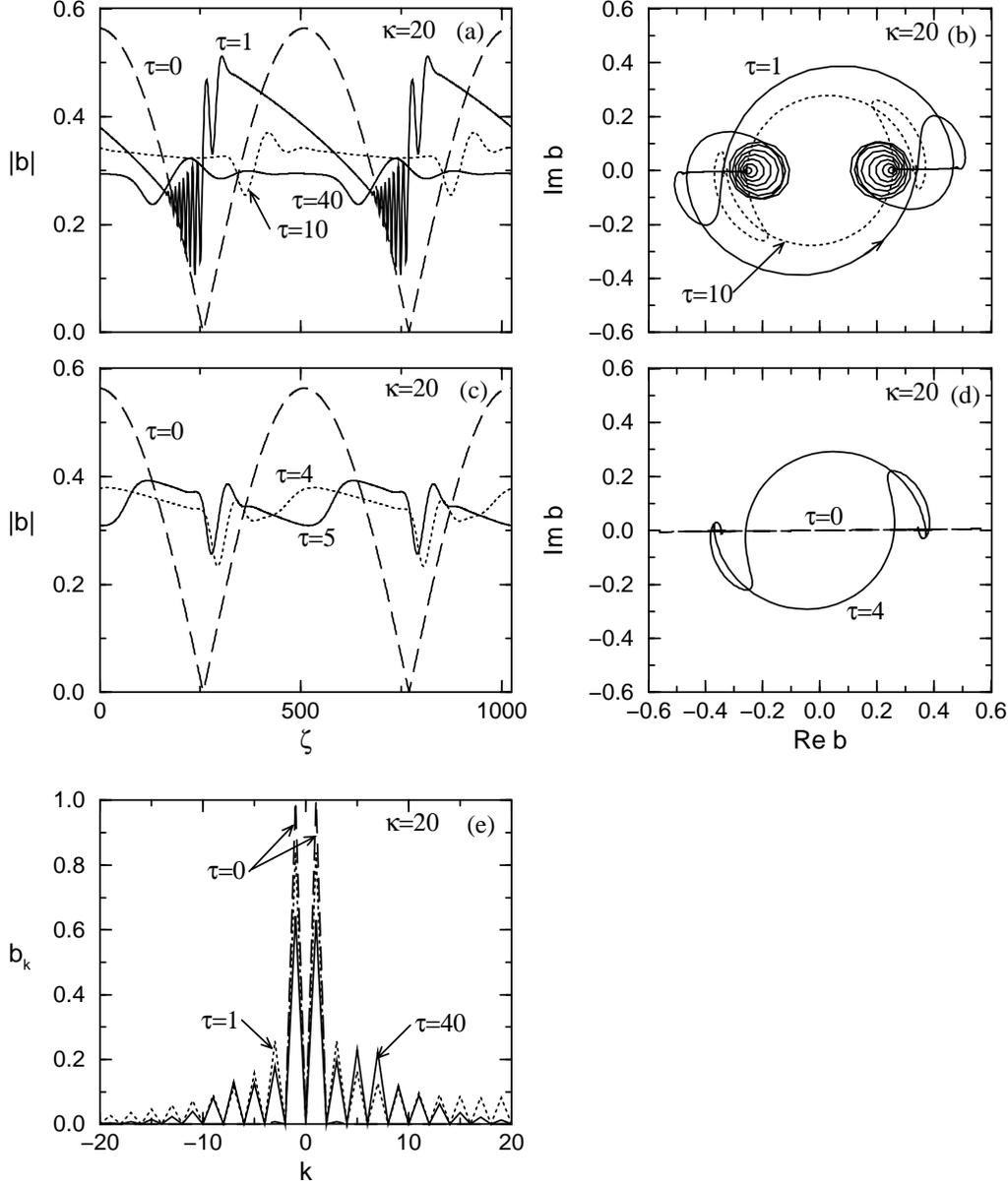,height=9.in}
\vskip-4cm
\caption{Wave evolution for the KNLS with collisional damping with 
$\beta=0.99$ and $\kappa=20$ (intermediate collisional regime) for  parallel 
propagating, {\em linearly} polarized sinusoidal wave 
initial condition. (a) and (c) - wave profiles, (b) and (d) - hodographs, 
(e) - harmonic spectra. At $\tau=0$ only $k=1, -1$ harmonics are excited.
The width of a shock wave is wider (as controlled by collisions), however the
width of the rotational discontinuity is the same, i.e.
it is controlled by  the dispersion parameter, alone.}
\label{fig:visclin+20}
\end{center}
\end{figure}

\begin{figure}[p]
\begin{center}
\psfig{file=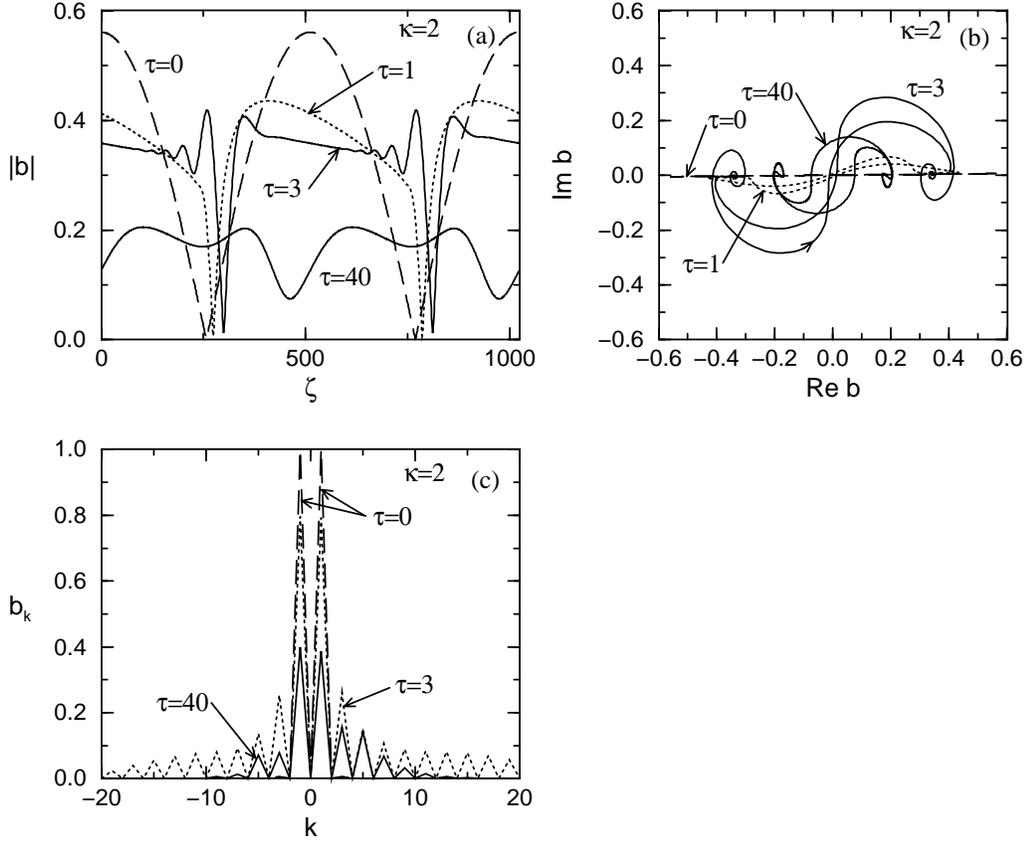,height=9.in}
\vskip-4cm
\caption{Wave evolution for the KNLS with collisional damping with 
$\beta=0.99$ and $\kappa=2$ (strongly collisional regime) for a parallel 
propagating, {\em linearly} polarized sinusoidal wave 
initial condition. (a) and (c) - wave profiles, (b) and (d) - hodographs, 
(e) - harmonic spectra. At $\tau=0$ only $k=1, -1$ harmonics are excited.
Note the remarkable similarity to the case of  Landau damping.}
\label{fig:visclin+2}
\end{center}
\end{figure}

\begin{figure}[p]
\begin{center}
\psfig{file=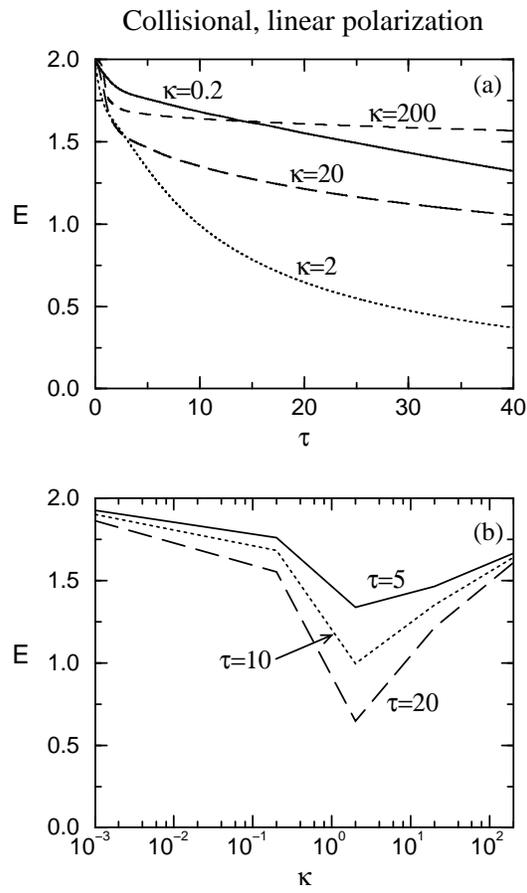,height=9.in}
\vskip-4cm
\caption{Wave energy evolution: (a) - temporal evolution of wave energy
im different collisional regimes. (b) - wave energy content at times 
$\tau=5, 10, 20$ as a function of plasma collisionality, $\kappa$.}
\label{fig:visclinenerg+}
\end{center}
\end{figure}

The nonlinear evolution of an initially linear polarization (in $\beta=0.99$
plasma) is shown in Figs.\ \ref{fig:visclin+200}, \ref{fig:visclin+20}, 
\ref{fig:visclin+2} for weakly collisional ($\kappa=200$), intermediate
($\kappa=20$), and strongly collisional ($\kappa=2$) regimes, respectively.
Energy dissipation is represented in Figs.\ \ref{fig:visclinenerg+}.
In a weakly collisional regime (Figs.\ \ref{fig:visclin+200}), a wave experiences
steepening of the back front, as a result of nonlinear dispersion set by the sign 
of $\kappa$ (or $L$) which, in turn, depends on plasma $\beta$. Since the 
nonlinear dispersion is positive in this case, it competes with the linear 
(negative) dispersion. The nonlinear dispersion is most 
important in regions of a high wave magnetic field (since it is proportional
to $|b|^3$), while conventional linear dispersion [last term in Eq.\ (\ref{mainKNLS})]
is important where $|b|$ is small. Parasitic small-scale circularly polarized
oscillations appear on the steep fronts when wave steepening is limited by 
dispersion (Figs.\ \ref{fig:visclin+200}a,b). These parasitic harmonics, 
however, disappear quickly as a consequence of the fact that high-$k$ 
components are strongly coupled to dissipation. The magnetic field vector 
rotates rapidly at the steepened edge of the wave. After the phase of nonlinear 
steepening, a propagating shock wave separates from the (standing) phase 
rotation region (i.e. the rotational discontinuity). The width of the shock front 
is controlled by collision length $L$ and its amplitude gradually decreases 
due to damping (Fig.\ \ref{fig:visclin+200}c). Such a process is remarkably 
similar to that first discussed by Cohen and Kulsrud \cite{Cohen}. However, 
it is just one limiting case ($\kappa\to\infty$) of the KNLS wave evolution. 
The spatial width of the rotational discontinuities is controlled by dispersion.
Hodographs of rotational discontinuities presented in Fig.\ 
\ref{fig:visclin+200}d are different from the Cohen-Kulsrud case. Here 
the magnetic field vector $b$ makes a counterclockwise rotation through
some angle (typically $50^\circ-90^\circ$) and 
than makes a clockwise rotation to complete a total $180^\circ$ phase jump
across the discontinuity. As shown in Fig.\ \ref{fig:visclin+200}e, harmonic 
spectra are broad, as collisional damping enters at small scales $k\ge\kappa$. 

Figs.\ \ref{fig:visclin+20} correspond to intermediate collisional damping 
($\kappa=20$). Parasitic small-scale, small amplitude oscillations are still 
emitted from the seepened front (Fig.\ \ref{fig:visclin+20}a,b). Propagating 
shocks  (Fig.\ \ref{fig:visclin+20}c) are sufficiently broad and disappear at a
higher rate than those in the weakly collisional case. 
Rotational discontinuities are localized at the regions of 
significant amplitude $|b|$ variations, and the typical hodograph takes on an
S-shape (however, with ``inflated'', almost semicircular central part). The 
harmonic spectrum
is narrower than in the $\kappa=200$ case (Fig.\ \ref{fig:visclin+20}e).
The rotational discontinuities move slowly to the right (as with 
Landau dissipation). Neither the sign of $\kappa$, nor the direction of shock
wave propagation affect the propagation direction of these phase 
discontinuities. More details of this case are shown in Figs.\ 
\ref{fig:visclin+20=15} at $\tau=15$. A remarkable similarity of these figures 
to those of Figs.\ \ref{fig:landlin=15} (Landau damping case) indicates a lack 
of a characteristic 
dissipation scale in the system, i.e. the dissipation scale is comparable with a
typical wave-length, $\kappa\sim k$.

The strong collisional regime is represented in Figs.\ \ref{fig:visclin+2}. 
This case exactly corresponds to that with Landau damping, since collisional
and wave lengths are equal (note, the $k$-scaling of damping is different).
No small-scale harmonics emerge, and shock wave generation is suppressed by
damping (Fig.\ \ref{fig:visclin+2}a), as typical of Landau dissipation S-shaped
hodograph displays (Fig.\ \ref{fig:visclin+2}b). Most of the energy is located 
in a few low-$k$ harmonics (Fig.\ \ref{fig:visclin+2}c). For a more strongly
collisional regime ($\kappa=0.2$, results are not shown) the effect of dissipation
becomes weaker, and wave evolution becomes similar to that of the weakly 
damped wave case. This effect is associated with the weakness of the resonant 
coupling (defined by $Q_2$ coefficient) of Alfv\'en and ion-acoustic waves.

The temporal evolution of wave energy is shown in Fig.\ \ref{fig:visclinenerg+}a
for several collisional regimes, from weak collisionality ($\kappa=200$) to a
collision dominated plasma ($\kappa=0.2$). Fig.\ \ref{fig:visclinenerg+}b
depicts total wave  energy content as a function of $\kappa$ [i.e. heat 
conduction coefficient as $(\chi_\|^c)^{-1}$] at different times $\tau=5, 10,
20$. The energy dissipation rate increases with collisionality (i.e. with
decrease of $\kappa$) until $\kappa\sim1$ and decreases for $\kappa<1$.
The last effect is due to the weak coupling of Alfv\'en and ion-acoustic
waves (the coefficient $Q_2$), which occures when the ion-acoustic frequency 
becomes smaller than its collisional damping rate.

\begin{figure}[p]
\begin{center}
\psfig{file=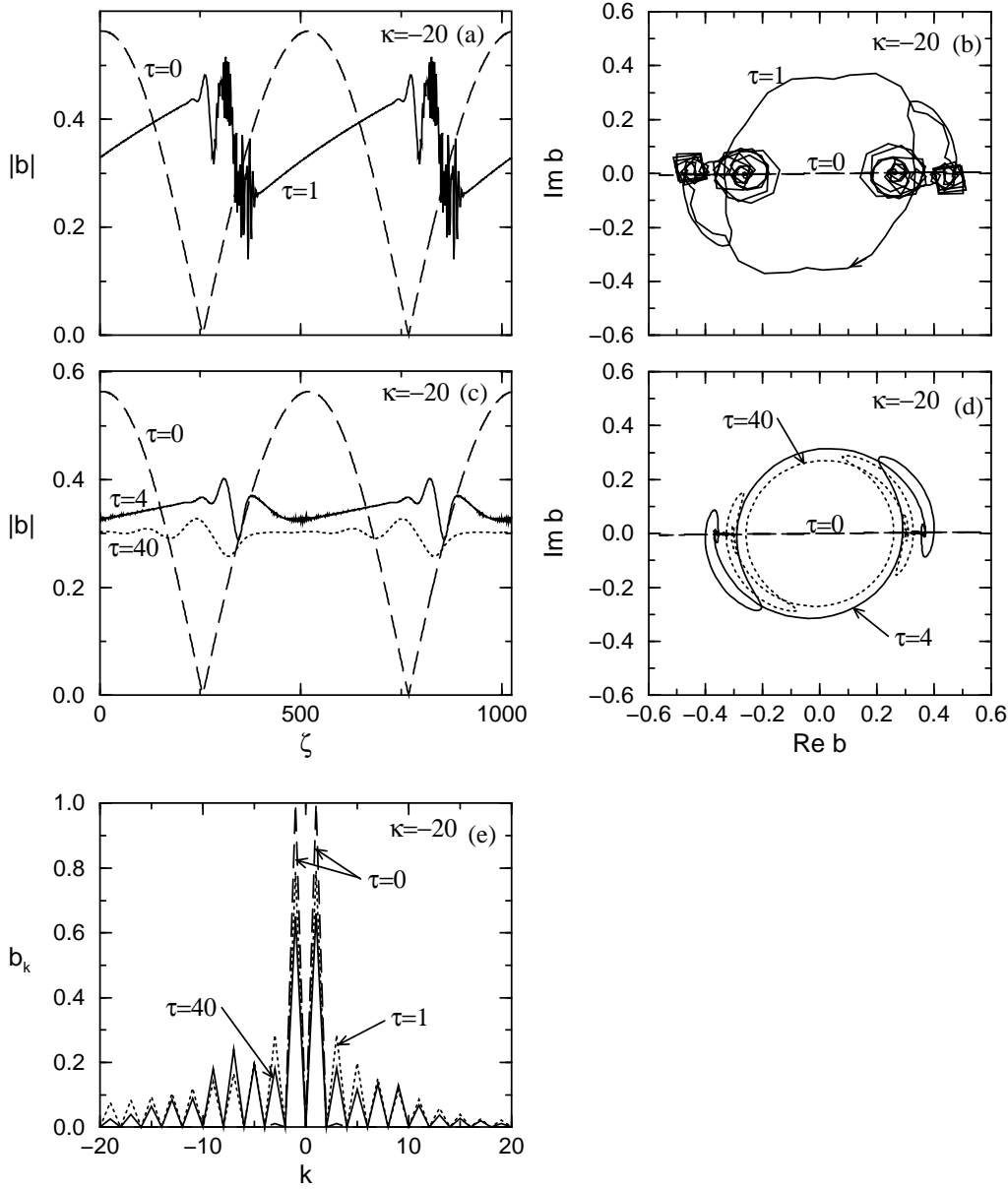,height=9.in}
\vskip-4cm
\caption{Wave evolution for the KNLS with collisional damping for
$\beta=1.01$ and $\kappa=-20$ (intermediate collisional regime) for a 
parallel propagating, {\em linearly} polarized sinusoidal wave 
initial condition. (a) and (c) - wave profiles, (b) and (d) - hodographs, 
(e) - harmonic spectra. At $\tau=0$ only $k=1, -1$ harmonics are excited.
Wavefront steepening occurs at the front face, thus it is controlled by the
sign of $\kappa$ (the nonlinear dispersion effect).}
\label{fig:visclin-20}
\end{center}
\end{figure}

Since the sign of $\kappa$ (or $L$) defines the sign of nonlinear dispersion, we
also considered the case of $\beta>1$. Traces of typical evolution are depicted 
in Fig.\ \ref{fig:visclin-20} for $\kappa=-20$ and $\beta=1.01$.
In this case both linear and nonlinear dispersions are negative
(and do not compete), so many small scale oscillations appear
at the (sharp) front of a wave train (Fig.\ \ref{fig:visclin-20}a,b). This `` noisy'' 
harmonic generation is possibly related to the modulation instability of 
steep fronts against small-scale modulations \cite{us}. The instability growth rate
in the collisionless regime decreases with increasing wave amplitude. In the 
collisional regime, the growth rate is amplitude independent. Thus, the large
amplitude suppression which is the case for collisionless damping is 
absent for diffusively dissipative Alfv\'en waves. Further evolution of
Alfv\'en trains generally follows the previous cases where $\beta<1$.

\begin{figure}[p]
\begin{center}
\psfig{file=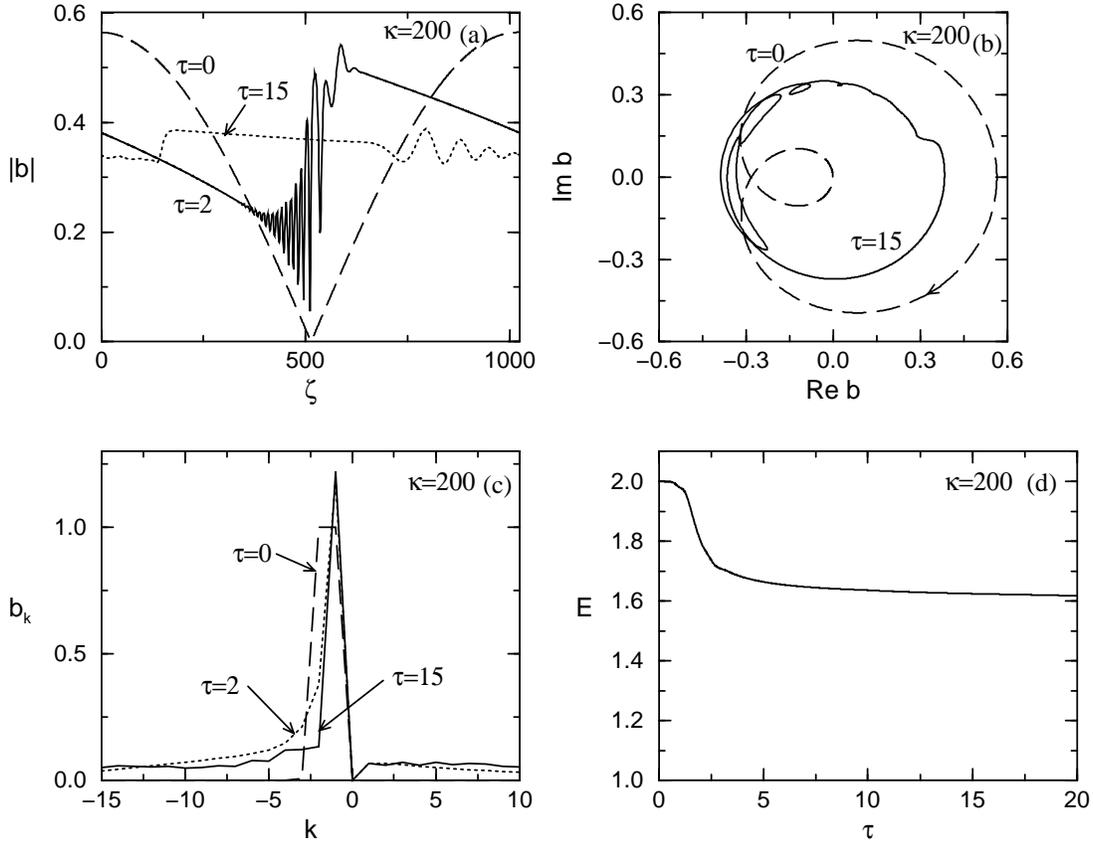,height=9.in}
\vskip-4cm
\caption{Wave evolution for the KNLS with collisional damping for
$\beta=0.99$ and $\kappa=200$ (weakly collisional regime) for a parallel 
propagating, {\em circularly} polarized sinusoidal wave initial condition. 
(a) - wave profiles, (b) - hodographs, (c) - harmonic spectra, 
(d) - temporal evolution of wave energy. At $\tau=0$ only 
$k=-1, -2$ harmonics are excited. A sharp front (shock) still forms,  but phase
irregularities form instead of a rotational discontinuity.}
\label{fig:visccirc+200}
\end{center}
\end{figure}

Circular polarizations evolve similarly to the Landau damping case. Despite
the emergence
of a shock wave (especially in a weakly collisional plasma), they do not 
evolve to rotational discontinuities. Instead, they generate smooth waveforms
with irregular phase rotation. The sign of $\kappa$  again defines the sign of
dispersion. Its value controls spatial scales. The result of dissipative
evolution is a dominant (lowest-$k$), circular harmonic and a low-amplitude
harmonic background tail, spreading up to scales $k\sim\kappa$. 
For illustration, Figs.\ \ref{fig:visccirc+200} show evolution of a circularly
polarized wave in a weakly dissipative plasma (more typical for astrophysical
situations) with $\beta=0.99$ and  $\kappa=200$.

\subsection{Discussion}

A significant difference between collisional (diffusive) and collisionless
(Landau) dissipation is the existence of a characteristic collisional scale
for the former case. As usual, it represents 
a ``boundary'' between the weakly and the strongly damped 
(i.e. low and high-$k$, respectively) parts of wave spectra. Integrating
Eq.\ (\ref{mainKNLS}) with $U_{NL}$ from Eq.\ (\ref{u-viscous}) over space
(see Section \ref{sec:landau}), we obtain:
\begin{eqnarray}
{\partial E\over\partial\tau}&=&-Q_2\int_{-\infty}^\infty
\left(\widehat{\cal J}_L\left[|b|^2\right]\frac{\partial}{\partial\zeta}|b|^2
+2|b|^2\frac{\partial}{\partial\zeta}\widehat{\cal J}_L\left[|b|^2\right] 
\right){\rm d}\zeta
\nonumber\\
&=&\frac{Q_2}{2}\int_{-\infty}^\infty 
\left|\left(|b|^2\right)_k\right|^2 
\frac{ik}{L^{-1}-ik}~ {\rm d}k
\nonumber\\
&=&Q_2\int_{0}^\infty 
\left|\left(|b|^2\right)_k\right|^2\frac{k^2}{\kappa^2+k^2}~ {\rm d}k~ .
\label{dE-viscous}
\end{eqnarray}
Thus, energy dissipation is controlled by both the coefficient $Q_2<0$
(which, in turn, controls the coupling of the Alfv\'enic mode to the damped
 ion-acoustic mode), and the dissipation scale 
$\kappa\propto(1-\beta)/\chi_\|^c$. Unlike the collisionless case, diffusion
enters at particular scales $k\ge\kappa$. As a result, shock waves may form
and their steepening will be limited by the collisional scale $\kappa$. 
The sign of $\kappa$ defines the sign of nonlinear dispersion (e.g. steepening
direction, etc.) and does not affect the dissipation rate. Since most  spectral
energy is associated with sufficiently small-$k$ components, the integral in
Eq.\ (\ref{dE-viscous}) increases with decreasing $\kappa$ until $\kappa$
is about unity. For smaller $\kappa$ the integral is approximately constant.
The coupling coefficient $Q_2$ decreases with decreasing $\kappa$
(for constant  $\beta$). The  higher the plasma collisionality, the weaker the
resonance of Alfv\'en and ion-acoustic waves, and thus the smaller $Q_2$ is.
Thus, the energy dissipation rate is {\em maximum}  when the dissipation
length is comparable with a typical wave-length. For a strongly collisional
regime, ion-acoustic wave  damping rate becomes larger than its frequency, 
so that ion-acoustic perturbations are no longer a plasma eigenmode.
Resonant coupling and energy transfer from an Alfv\'enic mode to
ion-acoustic oscillations is weak, despite strong dissipation of the latter.

One may mention the interesting case of strong damping 
($\kappa\simeq\lambda^{-1}_{wave}$). In this case, the system evolves as
if subject to collisionless dissipation, i.e. S-polarized KNLS dissipative
structures emerge from linear polarizations and circular polarizations quickly evolve
to a single harmonic state. However, such a scale-free regime
may occur only in strongly collisional plasma,
$\chi_\|^c\ge v_A^2/\Omega_i$, with $\beta$ very close to unity
($0.95\le\beta\le1.05$). Inspite of its rare occurrence in nature, this
case allows us to understand physical the reasons underlying the uniqueness of 
collisionless (Landau) dissipation for Alfv\'en wave dynamics.

The weakly collisional regime ($\kappa\to\infty$) seems more typical for
astrophysical and space plasmas. Nonlinear steepening of a wave front 
is first limited by nonlinear dispersion,  giving rise to radiation of 
small-scale, circularly polarized, noisy parasitic oscillations by the 
steep front (probably due to modulational instability). These oscillations 
damp quickly. At later times, the front separates into 
a shock wave moving in the direction of primary wave steepening and 
a waveform the type of which (e.g. directional or rotational discontinuity,
small-scale phase modulations, etc.)  depends on the initial wave polarization.
For initially linear polarizations, this waveform is a rotational discontinuity
(with a weak amplitude modulation imposed on it). For
circular polarizations, it is an intermittent, wavy structure
(the amplitude of perturbations of $|b|$ decreases with time), carrying weak, 
roughly arc-polarized phase irregularities. 
The rotational discontinuity (for $\beta<1$) 
consists of partially counterclockwise rotation through some angle (typically
$50^\circ-90^\circ$) followed by clockwise rotation through an angle larger than
$180^\circ$ to complete a $180^\circ$ phase difference across the
discontinuity. These waveforms have a characteristic spatial scale set by dispersion. 
The propagating shock wave front is much narrower and controlled by 
collisions. The amplitude of a shock decreases with time due to collisional
(diffusive) dissipation. The residual waveforms are weakly coupled to
nonlinear dissipation because $|b|\simeq const$ across them. In general,
linear polarizations are more strongly damped than circular ones.

Given that both linear and circular polarizations are initially excited, the 
weakly collisional quasi-stationary turbulence will consist of smooth
waveforms associated with rotational discontinuities and purely circularly
polarized, high amplitude waves.

Consider again Eq.\ (\ref{dE-viscous}) in a weak collisional regime, i.e. 
$\chi^c_\|\to0$. Then, as $\kappa\propto(1-\beta)/\chi^c_\|$,~ $\kappa$
is extremely large (and dissipation is very weak) for all values of $\beta$, except
$\beta\cong1$. The coefficient $Q_2\propto\beta/\chi^c_\|$ is not singular
at $\beta=1$. Thus, for $|\beta-1|\not\simeq0$ the dissipation rate
$\dot E\sim Q_2/\kappa^2\sim\chi^c_\|/(1-\beta)^2\to0$, indicating weak
wave damping. However, for $\beta$ close to unity, we infer from Eq.\
(\ref{dE-viscous}) that $\kappa\to0$ and $\dot E\sim Q_2\sim1/\chi^c_\|$
which is {\em larger}, for {\em weaker} collisionality. In space 
plasma, $\beta$ is a typically decreasing function of the distance from the sun
[i.e. $\beta=\beta(R)$], starting from $\beta>1$ near the sun and 
decreasing to $\beta<1$ further from it. Moreover, this plasma is only weakly
collisional. According to the discussion above, nonlinear waves propagating
out of the sun are extremely strongly damped in a narrow region at the 
distance $R_d$ where $\beta(R_d)\simeq1$. 
This region should be easily identifiable 
by a (much) higher (electron) temperature compared to elsewhere. 
The width of the region of enhanced dissipation may be estimated from the
condition $\kappa\propto\left[75\left(1-\beta(R)\right)/\chi^c_\|\right]
\left(v_A^2/\Omega_i\right)\simeq1$. Thus, the smaller the collisionality,
the thinner the high temperature region. 
Because of strong damping at $\beta=1$, waves
existing at distances either $R>R_d$ or $R<R_d$ never transit through the
point $R=R_d$ which is thus can act as a ``sharp boundary'' between these two
regions. Thus, the level of magnetic field perturbations will be low beyond
the point $R_d$ (assuming that the Alfv\'enic fluctiations have been generated 
near the sun), provided the width of the dissipation region is large enough.

\section{Gyro-kinetic (finite Larmor radius) effects on slightly 
oblique Alfv\'en trains}
\label{sec:larmor}
\subsection{Numerical solutions}

The evolution equation in this case is again Eq.\ (\ref{mainKNLS}). 
The nonlinear
velocity perturbation reads
\begin{eqnarray}
U_{NL}&=&\frac{v_A}{2}\Biggl\{\left[\left(M_1-\frac{N_1}{\Lambda^2}
\right)+\left(M_2-\frac{N_2}{\Lambda^2}\right)\widehat{\cal L}\right]
\label{u-flr}\\
& &{ }\!\!\!\!\!\!\!\cdot
\frac{\left(\widehat{\cal J}_\Lambda-
\widehat{\cal J}_{-\Lambda}\right) }{2\Lambda}
+\frac{1}{\Lambda^2}\left(N_1+N_2\widehat{\cal L}\right)\Biggr\}
\left[|b|^2-\langle|b|^2\rangle\right] ,
\nonumber
\end{eqnarray}
where
\begin{eqnarray*}
\Lambda^2&=&-2\eta^2_\|\beta{(1-\beta)+
[\widehat\chi^2_\|/\gamma](1-\beta/\gamma)\over
(1-\beta)^2+\widehat\chi^2_\|(1-\beta/\gamma)^2 } ,\\
N_1&=&\eta^2_\|{(1-2\beta)+\widehat\chi^2_\|(1-2\beta/
\gamma)\over(1-\beta)^2+\widehat\chi^2_\|(1-\beta/\gamma)^2 } ,\\
N_2&=&-2\eta^2_\|\widehat\chi_\|\beta{\gamma-1\over\gamma}
{1\over (1-\beta)^2+\widehat\chi^2_\|(1-\beta/\gamma)^2 } ,
\end{eqnarray*}
$\eta^2_\|\equiv(\rho_i^2/2)\sin^2{\Theta}$, and $\rho_i$ is the ion
gyro-radius. [Eq.\ (49) of Ref.\ \cite{us} contains a typographical error, in that
the two terms in round brackets were omitted]. Note here that $\Lambda^2$
can take on both positive and negative values. The operator $\widehat{\cal L}$
represents the  effects of resonant particles. The operator which contains Larmor
averaging is a composition of two collisional operators $\widehat{\cal J}_L$.
It can be written as 
\begin{equation}
\frac{1}{2\Lambda}\left(\widehat{\cal J}_\Lambda-
\widehat{\cal J}_{-\Lambda}\right)\doteq\frac{1}{2}
\left(\frac{\kappa}{\kappa-ik}+\frac{\kappa}{\kappa+ik}\right) ,
\label{oper-flr}
\end{equation}
where we redefine $\kappa=\bar z/\Lambda$. Thus it
can assume either real or imaginary values. 

In the limits $k\rho_i\ll1$ and $\Theta\ll1$ (i.e. $k/\kappa\ll1$), we
may expand denominators in the right hand side and plug into 
Eq.\ (\ref{u-flr}). Then, to lowest order, we write
\begin{eqnarray}
U_{NL}\!\Bigm|_{\Lambda\to0}&=&\frac{v_A}{2}
\Biggl\{\left(M_1+M_2\widehat{\cal L}\right)+
\left[\left(\Lambda^2M_1-N_1\right)\right.
\nonumber\\
& &\!\!\!\!\!\left.+\left(\Lambda^2M_2-N_2\right)
\widehat{\cal L}\right]\frac{\partial^2}{\partial\zeta^2}\Biggr\}
\left[|b|^2-\langle|b|^2\rangle\right] .
\end{eqnarray}
First, one may notice that there is no sharp transition from the $\Lambda^2>0$
regime
to the $\Lambda^2<0$ regime. Second, gyro-averaging the wave fields results
in nonlinear dispersion  of waves. It does not introduce additional dissipation.
It may, however, generate (due to modulational instability) 
a large number of  strongly damped high-$k$ harmonics.
Thus, it may affect the wave damping process.

\begin{figure}[p]
\begin{center}
\psfig{file=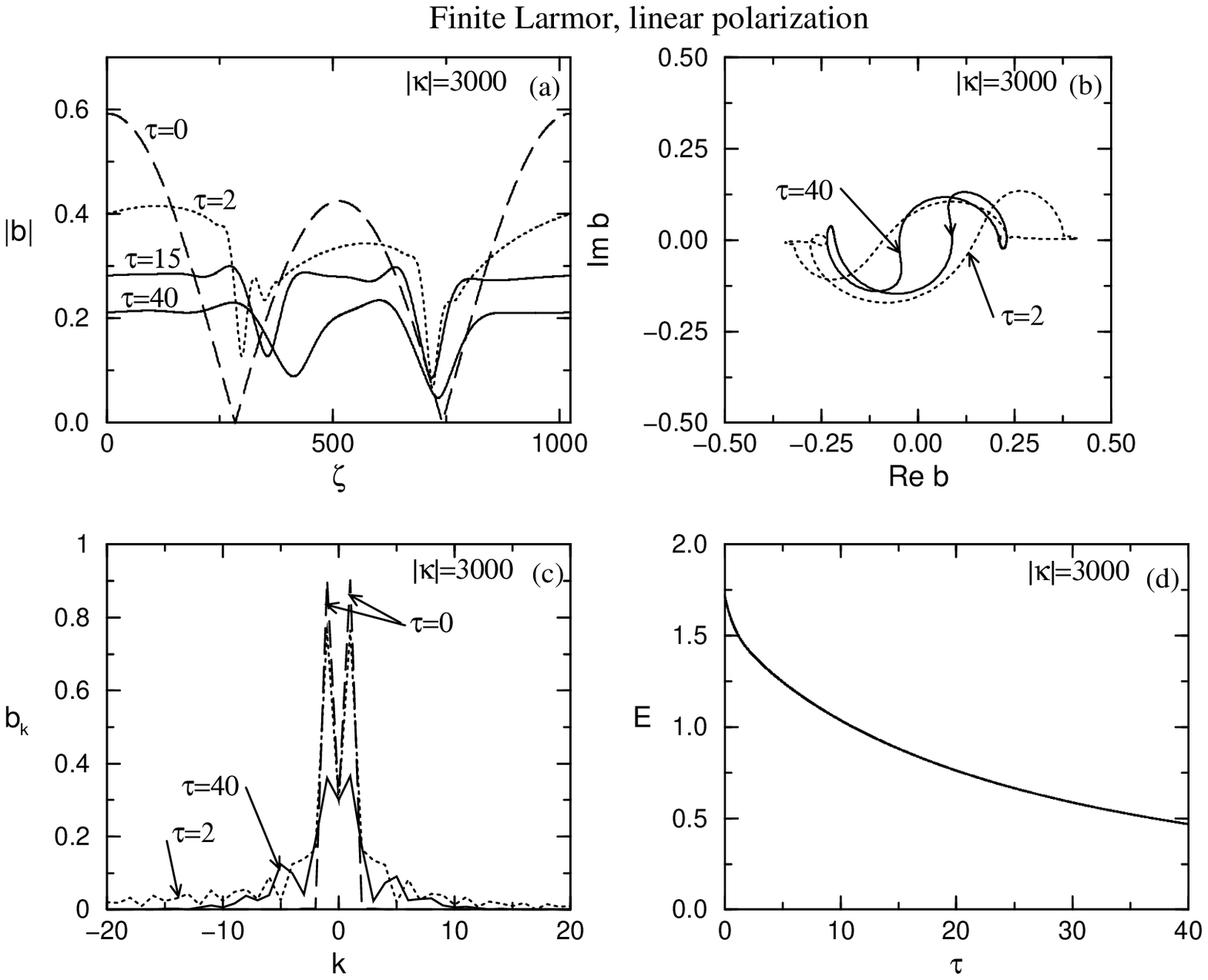,height=9.in}
\vskip-4cm
\caption{Wave evolution for the KNLS with Landau damping and
gyro-kinetic effects for $\kappa^2<0$, $|\kappa|=3000$ 
(typical $k\rho_i\simeq0.002$), and 
$\beta=0.9$ for an {\em obliquely} propagating
($\Theta=30^\circ$), {\em linearly} polarized sinusoidal wave 
initial condition. (a) - wave profiles, (b) - hodographs, 
(c) - harmonic spectra, (d) - temporal evolution of wave energy. 
At $\tau=0$ only $k=1, -1$ harmonics are excited. Observe that S-type
rotational/directional discontinuities form.}
\label{fig:flrlin3000-}
\end{center}
\end{figure}

\begin{figure}[p]
\begin{center}
\psfig{file=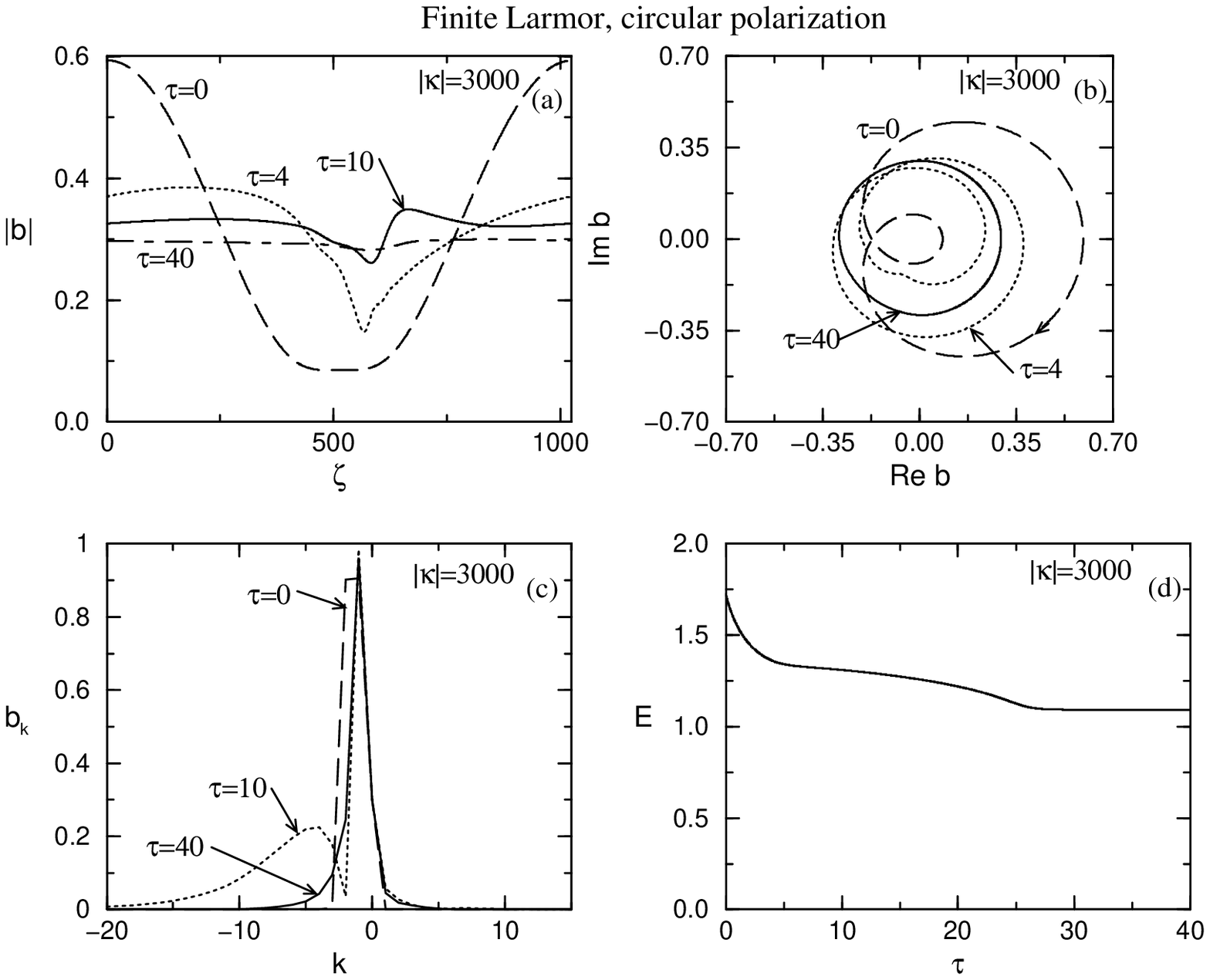,height=9.in}
\vskip-4cm
\caption{Wave evolution for the KNLS with Landau damping and
gyro-kinetic effects for $\kappa^2<0$, $|\kappa|=3000$
(typical $k\rho_i\simeq0.002$), and
$\beta=0.9$ for an {\em obliquely} propagating
($\Theta=30^\circ$), {\em circularly} polarized sinusoidal wave 
initial condition. (a) - wave profiles, (b) - hodographs, 
(c) - harmonic spectra, (d) - temporal evolution of wave energy. 
At $\tau=0$ only $k=-1, -2$ harmonics are excited. Note that arc-type
rotational discontinuities do not form.}
\label{fig:flrcirc3000-}
\end{center}
\end{figure}

Figs.\ \ref{fig:flrlin3000-} and \ref{fig:flrcirc3000-} depict the evolution of
linearly and circularly polarized initial conditions  (respectively) in
$\beta<1$ environment for $\kappa=3000$ which corresponds to 
$k\rho_i\simeq0.08$. The angle of wave propagation is $\Theta=45^\circ$
in both cases. There is no significant difference observed in the evolution of an
oblique linear polarization for zero and finite Larmor radii 
(compare Figs.\ \ref{fig:landofflinlong} and \ref{fig:flrlin3000-}, respectively).
Quasi-stationary KNLS dissipative structures (i.e. rotational discontinuities 
with $|b|^2\not= const$) form during the evolution. The width of each
is controlled by both linear and nonlinear (due to gyro-kinetic effects) 
dispersion. The harmonic spectrum, however, is more
irregular than in the case of zero Larmor radius (Fig.\ \ref{fig:flrlin3000-}c).
In contrast, an obliquely propagating,  circularly polarized wave behaves 
differently, and its
evolution is closer to the parallel propagating case. No 
arc-polarized structure forms from an initially circular polarization. Only
a waveform with dominantly low-$k$ circularly polarized harmonic 
forms instead (Fig.\ \ref{fig:flrcirc3000-}b).

\begin{figure}[p]
\begin{center}
\psfig{file=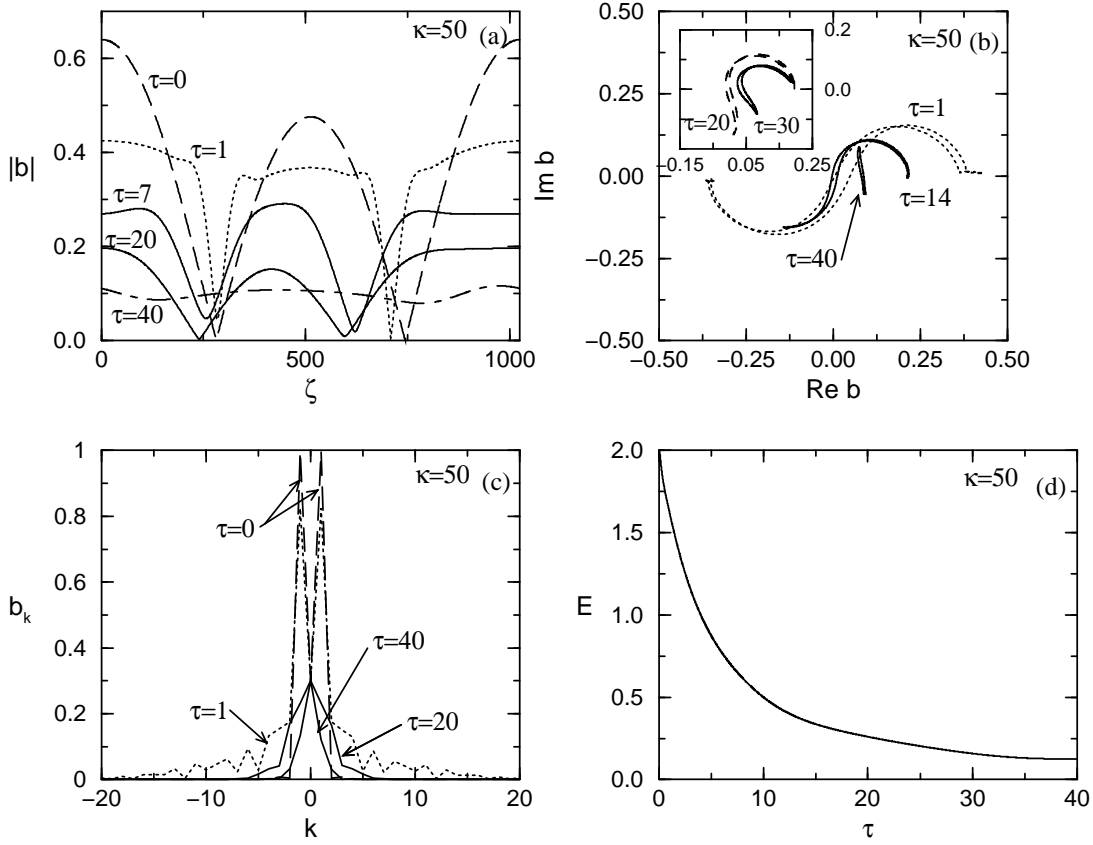,height=9.in}
\vskip-4cm
\caption{Wave evolution for the KNLS with Landau damping and
gyro-kinetic effects for $\kappa^2>0$, $|\kappa|=50$
(large Larmor radius, typical $k\rho_i\simeq0.1$), and
$\beta=1.3$ for an {\em obliquely} propagating
($\Theta=30^\circ$), {\em linearly} polarized sinusoidal wave 
initial condition. (a) - wave profiles, (b) - hodographs, 
(c) - harmonic spectra, (d) - temporal evolution of wave energy. 
At $\tau=0$ only $k=1, -1$ harmonics are excited.}
\label{fig:flrlin50+}
\end{center}
\end{figure}

Finite Larmor radius effects enter the energy dissipation rate, as shown in
Figs.\ \ref{fig:flrlin50+}. In this case $\beta=1.3, \kappa=50$ (i.e.
$k\rho_i\simeq0.5$), and $ \eta=0.2$. Again, two KNLS rotational
discontinuities of opposite senses of polarization form. They interact with 
each other quickly due to their increased width (set by dispersion), despite 
the fact that their
group velocities are similar. The combined effects of Landau damping and
nonlinear Larmor dispersion result in rapid wave energy decay.
This is shown in Figs.\ \ref{fig:flrlin50+}c,d.

We should mention another interesting case (results are not shown). For
$\beta<1$ and comparably small $|\kappa|$ ($\le50-100$), solutions of the
KNLS equation are unstable, i.e. the amplitudes of several harmonics increase 
rapidly with time. Our program does not allow us to follow the full dynamics
of such solutions, because a fast Fourier transform (used in the program)
works correctly only when harmonic amplitude vanishes quickly with increasing $k$. 
Note however that the perturbative derivation of the KNLS fails in this case, 
too. The physical understanding of this instability is discussed in the 
next subsection. 

\subsection{Discussion}

It is easily seen that finite Larmor radius corrections alone do not introduce a
damping process of any kind. Nevertheless, they can change the energy
dissipation rate via collisionless damping Indeed, it can even change the sign
of the dissipation,
as will be discussed later. Using our previous results, namely Eqs.\
(\ref{dE-landau}) and (\ref{dE-viscous}), we may now easily write
\begin{eqnarray}
{\partial E\over\partial\tau}&=&-\frac{1}{2}\int_{-\infty}^\infty
\left|\left(|b|^2\right)_k\right|^2 
\nonumber\\
& &{ }\cdot
\left\{\left(M_2-\frac{N_2}{\Lambda^2}\right)
\frac{\kappa^2}{\kappa^2+k^2}
+\frac{N_2}{\Lambda^2}\right\}|k|\ {\rm d}k .
\label{dEflr}
\end{eqnarray}
It is very important to emphasize that $\kappa^2$ is {\em not}
necessarily {\em positive}, but defined by the value of $\beta$ (i.e. 
$\kappa\propto1/\Lambda$). Thus, the last equation may be rewritten as
\begin{equation}
{\partial E\over\partial\tau}=\left\{
\begin{array}{rll}
&\displaystyle{\!\!\!\!\!
\int_{0}^\infty\left|\left(|b|^2\right)_k\right|^2
} \\ &\displaystyle{\!\!\!\cdot
\left(M_2\frac{\kappa^2}{\kappa^2+k^2}
+\frac{N_2}{\Lambda^2}\frac{k^2}{\kappa^2+k^2}\right)\! k~\!{\rm d}k},
&\kappa^2>0 ; 
\\[1.5em]
&\displaystyle{\!\!\!\!\!
\int_{0}^\infty\left|\left(|b|^2\right)_k\right|^2
} \\ &\displaystyle{\!\!\!\cdot
\left(M_2\frac{\left|\kappa^2\right|}{\left|\kappa^2\right|-k^2}
+\frac{N_2}{\left|\Lambda^2\right|}
\frac{k^2}{\left|\kappa^2\right|-k^2}\right)\! k~\!{\rm d}k},
\!&\kappa^2<0 .
\end{array}\right.
\eqnum{\ref{dEflr}$'$}
\label{dE-flr}
\end{equation}
Thus, nonlinear Larmor corrections for high-$\beta$ plasma (i.e.
$\kappa^2>0$) may result in enhanced dissipation, only. At low-$\beta$
(i.e. $\beta<1, \kappa^2<0$) they may also change the sign of
$\partial E/\partial\tau$ at least for $|k|>|\kappa|$, i.e. result in 
instability (i.e. pumping of a wave by plasma resonant particles). 
In the case of  $\beta$ less than unity, we replace $\kappa\to-i|\kappa|$
in Eq.\ (\ref{oper-flr}). It becomes
\begin{equation}
\frac{1}{2i\left|\Lambda\right|}
\left(\widehat{\cal J}_{i\left|\Lambda\right|}-
\widehat{\cal J}_{-i\left|\Lambda\right|}\right)\doteq\frac{1}{2}
\left(\frac{|\kappa|}{|\kappa|+k}+\frac{|\kappa|}{|\kappa|-k}\right) 
\end{equation}
which reveals singularities of the finite Larmor operator at $\pm k=|\kappa|$. 
The terms on the right hand side change their signs at $k=\mp |\kappa|$.
Thus, the sign of Landau damping ($\propto\widehat{\cal L}$) changes
sign for right hand polarized harmonics with $k<-|\kappa|$
(the first term) and left hand polarized waves with $k>|\kappa|$
(the second term). Most small-$k$
harmonics are damped on bulk plasma resonant particles. Since
plasma particles rotate over their Larmor circles (in the perpendicular plane)
along with longitudinal thermal motion, some large-$k$ harmonics
may experience growth  due to cyclotron resonance with the particles 
from the tail of the distribution function. 
Thus, finite Larmor radius effects together with the effect of resonant
particles in $\beta<1$ environment can drive an ion-cyclotron instability
(for growing harmonics $k>|\kappa|$) as well as ion-cyclotron damping
($k<|\kappa|$), enhanced dissipation 
(due to nonlinear dispersion) in $\beta>1$ plasma. Of course, the ion-cyclotron
instability saturates with time by, for example, particle distribution function
modification, an effect which is not included in this simple model.
For some values of plasma parameters, however, conventional Landau
damping may overcome generation of harmonics and suppress the instability.

\section{Conclusions}
\label{sec:concl}

In this paper, the influence of various kinetic effects (e.g. Landau and diffusive
dissipation as well as effects of the gyro-averaging of fluctuating fields
on a Larmor radius scale) were investigated in detail. A tractable analytic
evolution equation, the KNLS equation, was numerically solved to 
study nonlinear dynamics of finite-amplitude coherent Alfv\'en
waves in a finite-$\beta$,~ $T_i\simeq T_e$ plasma, natural 
to the solar wind. Previous DNLS studies failed to describe the evolution of the wide 
class of discontinuous waveforms which were recently observed in the solar wind.
These waveforms are the (i) {\em arc-polarized} rotational
discontinuities with phase difference {\em less} and {\em equal} to $\pi$ and
(ii) {\em S-polarized} rotational and directional discontinuities.
Current study shows that  {\em all} these discontinuous wave structures
discussed are {\em distinct solutions} of the same {\em simple} analytical
KNLS model for different initial conditions, e.g. initial wave polarization
and wave propagation angle. The results are summarized in Table\ \ref{table}.

{\bf I.}~ {\em Collisionless (Landau) damping} introduces a unique {\em 
scale invariant} dissipation process into the dynamics of Alfv\'enic trains. 
In contrast to viscous/diffusive dissipation, Landau damping is not restricted 
to small
scales. Thus, it is not surprisingly that the harmonic content of collisionless
dissipative waveforms tends to be dominated by large scales, with 
most wave energy contained in the low-$k$ harmonics. Since there is no 
characterictic dissipation scale in the system, the scale of the disssipative 
stuctures is set by {\em dispersion}, alone ({\em a l\`a} collisionless solitons and
shock waves). The dynamics
of Alfv\'en waves is sensitive to their sense of polarization, as well as
to their propagation direction with respect to the ambient magnetic field.
The principal results are:
\begin{enumerate}
\item {\em (quasi-), parallel initially linearly polarized waves} evolve rapidly
to form {\em dissipative structures} separated by regions of
linearly polarized Alfv\'en waves. These structures, which 
are naturally quasi-stationary, extend the traditional genre of collisionless
shocks to encompass the structures occurred via the competition of harmonic
generation (i.e. nonlinear wave steepening) and collisionless (Landau)
damping, rather than dispersion (i.e. as for solitons). Such structures are
commonly occurring over a wide range of plasma $\beta$, roughly:
$0.5\le\beta\le1.5$. These structures are characterized by fast phase changes
(phase jumps). Namely, the magnetic field vector rotates first through some
angle (typically, $50^\circ-90^\circ$), followed by rotation in the opposite
direction through an angle larger than $180^\circ$ to complete the total $\pi$
radian phase change. The phase diagram of the wave displays 
a distinct {\em S-shape}, which allows us to relate these dissipative
structures to a class of S-type directional/rotational discontinuities
observed in the solar wind. Unlike conventional  rotational discontinuities,
for which $|B|^2=const$, these KNLS rotational/directional 
discontinuities exhibit significant amplitude variation across the phase jumps.
The amplitude of these variations depends on the angle
of propagation and is largest for exactly parallel propagation.
The KNLS rotational/directional discontinuities have small velocity in the 
wave frame. Their directions are defined by the sign of the nonlinearity
coefficient $M_1$ (the discontinuity propagates with the wave for $M_2>0$).
The dissipative character of these structures is revealed by their increasing 
width and decreasing amplitude. Dissipation is strong in this case,
and energy decays roughly as $E(\tau)\propto1/(|M_2|\tau)$. Since the KNLS 
directional/rotational discontinuities display a prominent 
S-shaped phase portrait, they can (potentially) be regarded as a signature of strong 
Landau damping at work in satellite observations.
\item {\em (quasi-) parallel, initially circular polarizations} do not form
discontinuities. Since Landau damping preserves the initial symmetry of the
harmonic spectrum, circular polarizations form intermittent,
roughly arc-polarized waveforms which dissipate quickly, yielding
purely circularly polarized waves at the lowest $k$. Such waves
experience no nonlinear steepening and minimal damping. The different 
evolution of linear and circular polarizations is due to specific coupling of the 
waves to collisionless damping, which is, in turn, sensitive to the (initial)
symmetry of the harmonic spectrum.
\item  {\em (quasi-) parallel, initially elliptically polarized waves} represent
an intermediate case between linear and circular polarizations. Well
localized, rotational discontinuities (similar to dissipative structures)
form. They are separated by large regions of linearly polarized waves.
Amplitude variations across the discontinuities are  small.
These waveforms are more weakly coupled to dissipation. The total
energy of the waveforms decays as 
$E(\tau)\propto\exp(-\Delta|M_2|\tau)$, where 
$\Delta=|b_{+1}|^2-|b_{-1}|^2$ is the difference in energy between the two
initially excited harmonics.
\item {\em oblique, initially linearly polarized, longitudinal waves} form KNLS
rotational/directional discontinuities (dissipative structures) with opposite
phase rotations, having different group velocities. Thus, these wave
structures may interact to annihilate each other. 
As these waves are nonlinearly coupled to the (large) 
ambient field $B_\bot=B_0\sin\Theta$, they are connected to dissipation 
more strongly than quasi-parallel waves are.
Thus, the wave energy dissipate rapidly: $E(\tau)\propto\exp(-|M_2|\tau)$.
Residual, small-scale, weakly-nonlinear, weakly-damped Alfv\'enic 
perturbations may survive for a long time. 
\item {\em oblique, initially circularly polarized waves and elliptically polarized,
longitudinal waves} do not generate
dissipative structures. Despite this, they  evolve to stationary waveforms 
with negligible amplitude and large phase variations. Their phase diagram 
displays a pure arc, with the arc-angle less than $180^\circ$. Fast magnetic 
vector rotation along the arc at the discontinuity is followed by a part of an
oppositely rotating circularly polarized wave. Thus, these structures
have all requisites to be identified with {\em arc-polarized rotational
discontinuities} observed in the solar wind. This waveform experiences 
{\em no damping} and nonlinear steepening. However, the energy content 
of this waveform is not small.
\item {\em oblique,  transverse waves} evolve to stationary 
{\em arc-polarized rotational discontinuities} with the arc-angle less 
than $180^\circ$. The energy content of this waveform is large.
\end{enumerate}
According to the discussion above, nonlinear Alfv\'en waves subject to
damping on resonant particles evolve to narrow-spectrum waveforms
with most energy concentrated in lowest-$k$ harmonics.
The initial spectrum symmetry (i.e. the wave polarization) is generally 
preserved during the evolution. Based on the above, one can suggest that
(given initial equal populations of isotropically distributed circular and 
linear polarizations) quasi-parallel magnetic field fluctuations will
consist of predominantly circularly polarized waves and lower amplitude 
S-polarized KNLS directional discontinuities, while oblique perturbations
are predominantly arc-polarized discontinuities, separated by pieces of
oppositely circularly polarized waves. One should emphasize that {\em all}
wave structures emerge naturally through the nonlinear evolution of
simple, sinusoidally amplitude modulated Alfv\'en waves with no
{\em a priori} assumptions or special initial conditions used.

{\bf II.}~ The dynamics of nonlinear Alfv\'en wave trains subject to 
{\em diffusive (collisional) damping} differ significantly. There are two
different regimes of weak and strong plasma collisionality, controlled by
the ratio of dissipation length $L$ to the typical wavelength $\lambda$,
i.e. $|L|\ll\lambda$ and $|L|\ge\lambda$, respectively.
The results are:
\begin{enumerate}
\item The effect of collisions introduces nonlinear dispersion into the dynamics, 
along with a nonlinear damping process. Thus  it affects the
direction of steepening, rate of spectrum modification, etc..
\item The sign of $L$ defines the sign of {\em nonlinear dispersion}
which, in turn, is controlled by plasma $\beta$, i.e. $L\propto(1-\beta)$.
Thus, for $\beta<1$ and $\beta>1$, nonlinear steepening occurs
at the rear and front edges of a train, respectively.
\item The absolute value of $L$ affects the {\em dissipation rate}.
The damping rate has a maximum near $|L|\simeq\lambda$. In  the weakly 
collisional regime, the dissipation process is slower on account of infrequent 
collisions.
However in higher collisionality plasma, the acoustic mode is strongly
damped, so it is no longer a plasma eigenmode (i.e. $\omega<\gamma$,~
here $\gamma$ is the damping rate). Thus, resonant energy transfer 
from Alfv\'enic to acoustic waves is suppressed.
\item In a {\em weakly collisional regime}, a nonlinear wave evolves to a
propagating shock wave (the direction of propagation coincides with the
direction of wave steepening) and a rotational discontinuity or phase 
irregularities, depending on an initial polarization, i.e. whether the initial
polarization is linear or circular, respectively. The shock amplitude gradually
decreases due to damping. The thickness of the shock is controlled by
collisions, however the spatial size of the phase discontinuity/irregularities
is larger and controlled by dispersion. In a {\em strongly  
collisional regime}, the wave evolution is much similar to the case of
Landau damping. Obviously, the system lacks the characteristic 
dissipation scale in this limit, i.e. $|L|\sim\lambda$.
\item The dissipation process is very sensitive (especially in {\em weakly}
collisional plasma) to the value of $\beta$, and sharply peaked at $\beta$ 
equal to unity. If one assume $\beta=\beta(R)$ is a function of the distance
from the sun, then waves generated near the sun and propagating outward
will be strongly damped at the distance $R_d$ defined by $\beta(R_d)=1$.
One can, thus, expect increased plasma temperature in this region, as well as
decreased magnetic field fluctuation level  (i.e. characteristic amplitude 
of Alfv\'en waves) beyond this point, i.e. at distances $R>R_d$. This process
may be important, however, when the width of the dissipation region is large 
enough (comparable to or exceeding the typical wavelength). This is, in turn, controlled
by the spatial profile of $\beta$. 
\end{enumerate}

{\bf III.}~ Effects of {\em finite Larmor radius} impact the dynamics of
obliquely propagating waves, only. They are associated with small-scale
gyro-averaging  of fluctuating fields on a scale of Larmor orbits of plasma
particles. These effects 
\begin{enumerate}
\item introduce {\em nonlinear dispersion} controlled by the plasma $\beta$, 
Larmor scale $\rho_i$, and angle of propagation $\Theta$.
\item typically {\em suppress} the formation of arc-polarized discontinuities 
from initially circular polarizations, but do not change the qualitative evolution of 
linear polarizations.
\item affect the wave energy damping rate in high-$\beta$ plasma
($\beta>1$) and may even change the sign of the dissipation in low-$\beta$
plasma ($\beta<1$). The last case corresponds to an instability. Thus, as
a result of gyro-motion of resonant particles interacting with a wave packet 
(in $\beta<1$ plasma), an ion-cyclotron resonance occurs. It results in 
damping of low-$k$ harmonics and growth (instability) of high-$k$
harmonics, which interact with the tail of a particle distribution function.
Thus, small-scale bursts may be generated in this regime.
\end{enumerate}
A more detailed investigation of obliquely propagating waves will be presented 
in future publications. 

\subsection*{Acknowledgments}
We wish to thank B. T. Tsurutani, V. D. Shapiro, and S. K. Ride for useful 
discussions. This work was supported by Department of Energy Grant No.  
DE-FG03-88ER53275, Natinal Aeronautics and Space Administration 
Grants No.  NAGW-2418 and No. 10-85849, and National Science Foundation 
Grant No. ATM-9396158  (with the University of California, Irvine).

\onecolumn
\begin{table}[p]
\begin{center}
\psfig{file=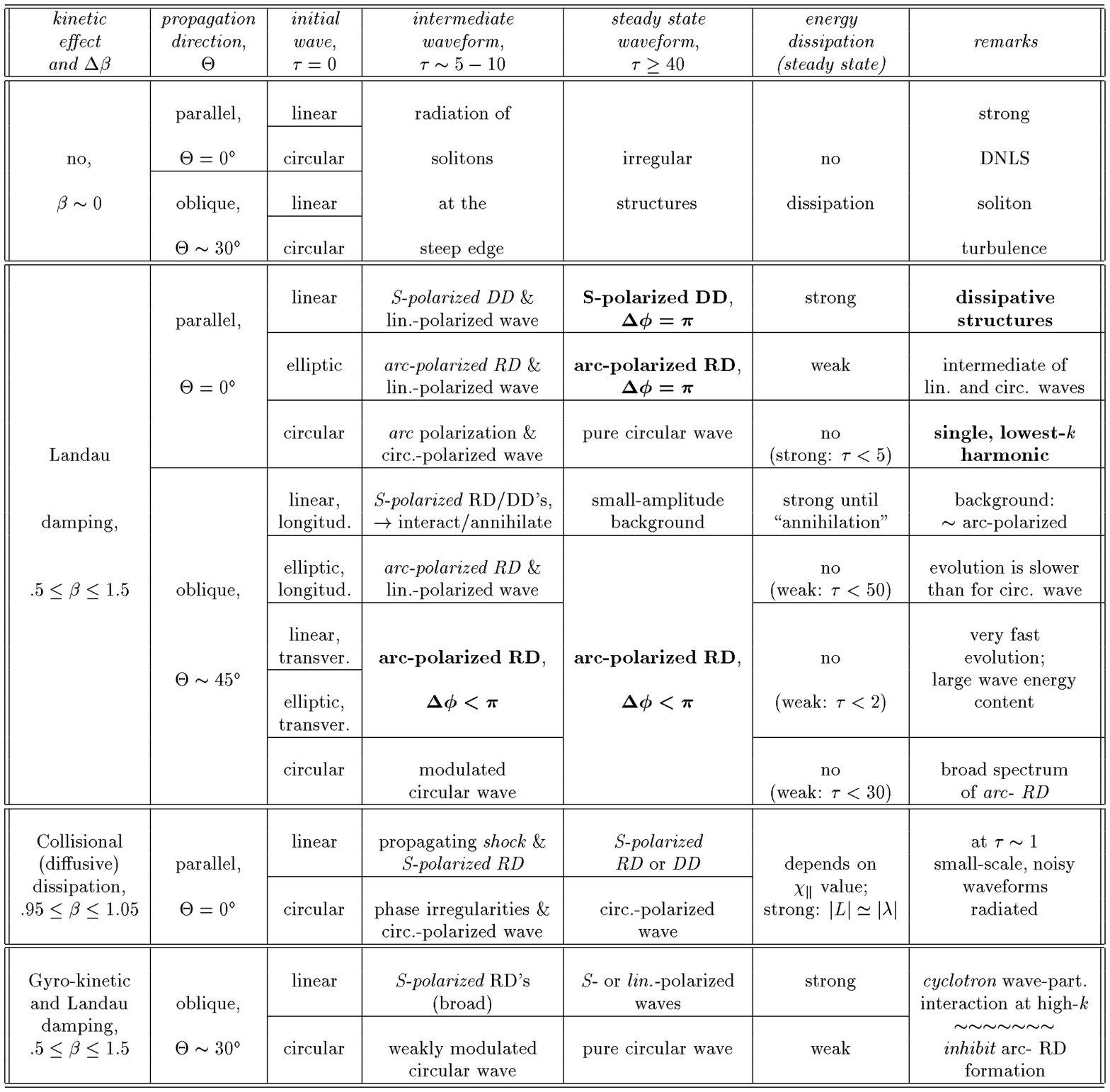,height=9.in}
\vskip-1cm
\caption{Summary of the numerical solutions.}
\label{table}
\end{center}
\end{table}
\end{document}